\documentclass[11pt,a4paper]{article}
\usepackage[T1]{fontenc}
\usepackage[utf8]{inputenc}
\usepackage{lmodern}
\usepackage[margin=25mm]{geometry}
\usepackage{amsmath,amssymb}
\usepackage{graphicx}
\usepackage{fixltx2e}
\makeatletter
\long\def\@makecaption#1#2{\vskip\abovecaptionskip
  {\small\noindent\textbf{#1.} #2\par}\vskip\belowcaptionskip}
\makeatother
\usepackage[hidelinks]{hyperref}

\begin{document}

{\LARGE\bfseries Factorization dynamics between quantum Fisher information and quantum coherence\par}

Xinzhi Zhao\textsuperscript{1}, Xinglei Yu\textsuperscript{1}, Liangsheng Li\textsuperscript{2}, Wenting Zhou\textsuperscript{1}, Chengjie Zhang\textsuperscript{1*}

\textsuperscript{1} School of Physical Science and Technology, Ningbo University, Ningbo 315211, China.

\textsuperscript{2} National Key Laboratory of Scattering and Radiation, Beijing 100854, China.

*Corresponding author. Email: cjzhang@ustc.edu

\section*{Abstract}

Quantum Fisher information (QFI) quantifies the sensitivity of a quantum state to a parameter change and plays a key role in quantum metrology. Meanwhile, quantum coherence is a crucial resource for quantum information processing. However, despite extensive studies on various aspects of quantum metrology, the interplay between the dynamics of the QFI and quantum coherence remains unexplored. Here, we explore the factorization relationship between the QFI and quantum coherence, both theoretically and experimentally. We prepare pure states in qubit and qutrit systems and apply unitary evolution to investigate this relationship. Our results confirm the factorization law, offering strong evidence for this relationship. Furthermore, we establish a lower bound on the variance of the estimated parameter that does not require the final state information. This approach reveals a deep connection between the QFI and quantum coherence, providing insights into quantum parameter estimation. These findings have potential applications in quantum sensing and quantum metrology.

\section*{INTRODUCTION}\label{sec1}

Quantum parameter estimation is a key topic in many scientific and technological fields, as it aims to find the unknown parameters of a quantum system with high precision. Phase estimation is a common problem in this area, which involves challenges such as state preparation, loss, and decoherence (\emph{\ref{R1}}--\emph{\ref{R6}}). The main tool for quantum parameter estimation is the quantum Fisher information (QFI) (\emph{\ref{R7}}--\emph{\ref{R12}}), which measures the sensitivity of a quantum state concerning changes in the parameters encoded in it, where its inverse is known as the quantum Cramér-Rao bound (QCRB) (\emph{\ref{R13}}, \emph{\ref{R14}}). The QFI has many applications in quantum information theory, such as uncertainty relations (\emph{\ref{R15}}, \emph{\ref{R16}}), entanglement detection (\emph{\ref{R17}}--\emph{\ref{R19}}), characterizing the non-Markovianity of open systems (\emph{\ref{R20}}--\emph{\ref{R22}}), and determining the optimal measurements for estimating parameters (\emph{\ref{R23}}--\emph{\ref{R25}}). The QFI is a versatile and powerful concept that has been studied extensively in various scenarios and settings (\emph{\ref{R26}}--\emph{\ref{R28}}).

As a central task in quantum metrology, quantum parameter estimation has brought renewed attention to quantum coherence, recognizing that quantum coherence plays a key role in enhancing precision (\emph{\ref{R29}}--\emph{\ref{R32}}). Quantum coherence is a crucial physical resource that enables interference phenomena and quantum information processing (\emph{\ref{R29}}--\emph{\ref{R51}}). Coherence was originally a concept for describing wave interference, but quantum coherence has become a central topic in quantum engineering, such as quantum metrology (\emph{\ref{R30}}--\emph{\ref{R32}}), quantum key distribution (\emph{\ref{R34}}), and quantum computing (\emph{\ref{R35}}). Recently, it has also been recognized for its crucial role in quantum secure direct communication (\emph{\ref{R36}}, \emph{\ref{R37}}) and high-order quantum Pancharatnam-Berry phase studies. Quantum coherence is the essence of quantum physics and its advantage over classical physics. Various methods have been proposed to detect and quantify quantum coherence on a fixed basis, such as coherence witnesses (\emph{\ref{R31}}, \emph{\ref{R32}}, \emph{\ref{R38}}), trace distance (\emph{\ref{R39}}, \emph{\ref{R40}}), and skew information (\emph{\ref{R41}}, \emph{\ref{R42}}), as well as other methods in a basis-independent way (\emph{\ref{R43}}--\emph{\ref{R48}}). In addition, recent works have highlighted the role of quantum coherence in various systems. For example, in the two-electron spin system, coherence and entanglement are influenced by interaction parameters and external magnetic fields within certain ranges (\emph{\ref{R52}}). Similarly, the analysis of the fractional Schrödinger equation has shown that the QFI and coherence can be related under fractional evolution (\emph{\ref{R53}}).

While quantum coherence serves as a key resource and the QFI quantifies the ultimate precision limit in quantum metrology, the precise relationship between them remains insufficiently understood. Furthermore, gaining a deeper understanding of this relationship is of great importance. Quantum coherence refers to the ability of a quantum system to maintain superposition. This property serves as a valuable resource in improving the performance of quantum metrology tasks, such as the phase discrimination game (\emph{\ref{R33}}) and quantum parameter estimation (\emph{\ref{R29}}, \emph{\ref{R34}}, \emph{\ref{R35}}). A simple example of quantum metrology where quantum coherence is essential was given in (\emph{\ref{R29}}, \emph{\ref{R34}}): Suppose we have a D-dimensional Hilbert space and a nondegenerate Hamiltonian $\hat{H}$, whose eigenvectors form the reference basis for quantifying quantum coherence. Then, an output state $\rho_{\theta} = U_{\theta}\rho U_{\theta}^{\dagger}$, where $U_{\theta} = e^{- i\hat{H}\theta}$, can be used to estimate an unknown parameter $\theta$ if and only if the input state $\rho$ has nonzero coherence on the reference basis. Moreover, as it is known, the relationship between the QFI and quantum coherence remains to be widely explored. There are still several intriguing open questions that have yet to be addressed: How does quantum coherence determine the QFI in general quantum systems? Is it possible to establish a lower bound on parameter estimation variance based solely on the coherence of the initial state when the final state information is unavailable? To address these challenges, our work bridges this gap by presenting a complete dynamical process linking the coherence of the input initial state to the QFI, which typically depends on the final state.

In this work, we theoretically and experimentally investigate the factorization law between the QFI and the coherence for pure states, which states that under unitary transformations, the QFI of an output state is bound by the product of the maximal QFI and the coherence of its input pure state. We test our theory in qubit and qutrit cases and find good agreement between the experimental and theoretical results. Our experimental results provide evidence for the factorization law between the QFI and quantum coherence. Furthermore, we conduct experiments on quantum parameter estimation under optimal measurement conditions in the two-dimensional (2D) scenario. This approach demonstrates that the QFI, central to quantifying precision limits in quantum metrology, can be directly inferred from the quantum coherence of the initial state, bypassing the need to analyze the final state. This highlights the potential of quantum coherence as a resource for enhancing estimation accuracy, as it fundamentally influences the precision limits in quantum metrology, independent of the final state measurement outcome.

The paper is organized as follows. The next section establishes the theoretical framework that supports our research, providing a foundation for the subsequent experimental design and analysis. Then, we introduce the experimental design in detail, including the setup, procedures, and the rationale behind our methodological choices. The results of the experiments are presented and analyzed thoroughly, highlighting the key findings and their implications. Last, we summarize the main contributions of this work and suggest potential directions for future research.

\section*{RESULTS}\label{sec2}

\subsection*{Connection between the QFI and quantum coherence}\label{sec3}

The QFI is a measure of the precision of a quantum measurement, while the QCRB is the limit of that precision. Here, the QCRB is expressed as $\text{Var}( \hat{\theta} ) \geq \lbrack n\mathcal{F}_{\theta} \rbrack^{- 1}$, where \emph{n} denotes the number of measurements, $\hat{\theta}$ denotes an unbiased estimator of $\theta$, and $\mathcal{F}(\theta)$ refers to the QFI. The variance of any unbiased estimator is at least greater than the inverse of the QFI. That is, the QFI defines the theoretical limiting boundary on the precision of the variance that can be achieved by the optimal parameter estimates (\emph{\ref{R9}}--\emph{\ref{R11}}, \emph{\ref{R23}}).

In this study, we consider one of the simplest forms of QFI, i.e., in the pure state, the Hamiltonian quantity $\hat{H}$ is independent of the parameter $\theta$. Then, the quantum state $| \psi_{\theta} \rangle$ with parametric information is $| \psi_{\theta} \rangle = U_{\theta} | \psi \rangle = e^{- i\hat{H}\theta} | \psi \rangle$, and the QFI with parameter information is the form of the following variance

\begin{equation}
\mathcal{F}_{\theta}(|\psi_{\theta}\rangle)=4\bigl[\langle\psi_{\theta}|\widehat{H}^{2}|\psi_{\theta}\rangle-\langle\psi_{\theta}|\widehat{H}|\psi_{\theta}\rangle^{2}\bigr]
\label{eq:1}
\end{equation}

The QFI is maximized when the input state is a superposition of eigenstates corresponding to maximal and minimal eigenvalues, a situation referred to as the channel QFI, which can be expressed as (\emph{\ref{R12}})

\begin{equation}
\mathcal{F}_{\theta,\mathrm{max}} = {(\lambda_{\max} - \lambda_{\min})}^{2}
\label{eq:2}
\end{equation}

$\lambda_{\text{max}}$ and $\lambda_{\text{min}}$ correspond to the maximum and minimum values of the eigenvalues of $\hat{H}$, respectively.

Coherence serves as a fundamental physical resource in quantum systems, so we can use it to get the expression form of the QFI that includes state coherence. Theoretically, the coherence of a quantum state can be measured by measuring its off-diagonal elements; the $l_{2}$-norm of coherence is defined as (\emph{\ref{R54}})

\begin{equation}
C_{l_{2}}(\rho) = \sum_{\begin{array}{r}
i,j \\
i \neq j
\end{array}}^{}{|\rho_{i,j}|}^{2}
\label{eq:3}
\end{equation}

For a pure state of one qubit ($D=2$), we obtain the factorization law of the QFI and coherence as

\begin{equation}
\mathcal{F}_{\theta}(|\psi_{\theta}\rangle) = 2C_{l_{2}}(|\psi\rangle)\mathcal{F}_{\theta,\mathrm{max}}
\label{eq:4}
\end{equation}

The detailed construction is described in Materials and Methods. Moreover, for a pure state of one qudit ($D>2$), the factorization law of the QFI and coherence can be obtained as

\begin{equation}
\mathcal{F}_{\theta}(|\psi_{\theta}\rangle) \leq 2C_{l_{2}}(|\psi\rangle)\mathcal{F}_{\theta,\mathrm{max}}
\label{eq:5}
\end{equation}

it should be noted here that the selection of the initial state must meet the condition of $\sum_{i < j}{| a_{i} |}^{2}{| a_{j} |}^{2} < 0.25$, and \hyperref[eq:5]{Eq. 5} becomes trivial when $\sum_{i < j}{| a_{i} |}^{2}{| a_{j} |}^{2} \geq 0.25$. The detailed construction is described in Materials and Methods. On the basis of the QCRB and \hyperref[eq:5]{Eq. 5}, we can also derive a lower bound for the variance $\text{Var}( \hat{\theta} )$ as (see the Supplementary Materials for details)

\begin{equation}
\operatorname{Var}\left( \widehat{\theta} \right) \geq \frac{1}{n\mathcal{F}_{\theta}(|\psi_{\theta}\rangle)} \geq \frac{1}{2nC_{l_{2}}(|\psi\rangle)\mathcal{F}_{\theta,\mathrm{max}}}
\label{eq:6}
\end{equation}

where \emph{n} denotes the number of measurements, $\hat{\theta}$ is said to be an unbiased estimator of $\theta$, and $\mathcal{F}_{\theta}(\theta)$ refers to the QFI of final states.

An illustration of the factorization law of the QFI and coherence is demonstrated in \hyperref[F1]{Fig. 1}. On the basis of \hyperref[eq:4]{Eqs. 4} and \hyperref[eq:5]{5}, one can conclude that if the coherence of an input state is zero {[}$C_{l_{2}}( | \psi \rangle ) = 0${]}, then the QFI of its output state must be zero {[}$\mathcal{F}_{\theta}( | \psi_{\theta} \rangle ) = 0${]}. In other words, the input state must be coherent such that its output state can be used for estimating an unknown parameter $\theta$. Moreover, \hyperref[eq:5]{Eq. 5} establishes the upper bound that the QFI can attain. In the following, we will experimentally verify the factorization law between the QFI and coherence by testing \hyperref[eq:4]{Eqs. 4} and \hyperref[eq:5]{5}.

\begin{figure}[t]
\centering
\includegraphics[width=0.62\linewidth]{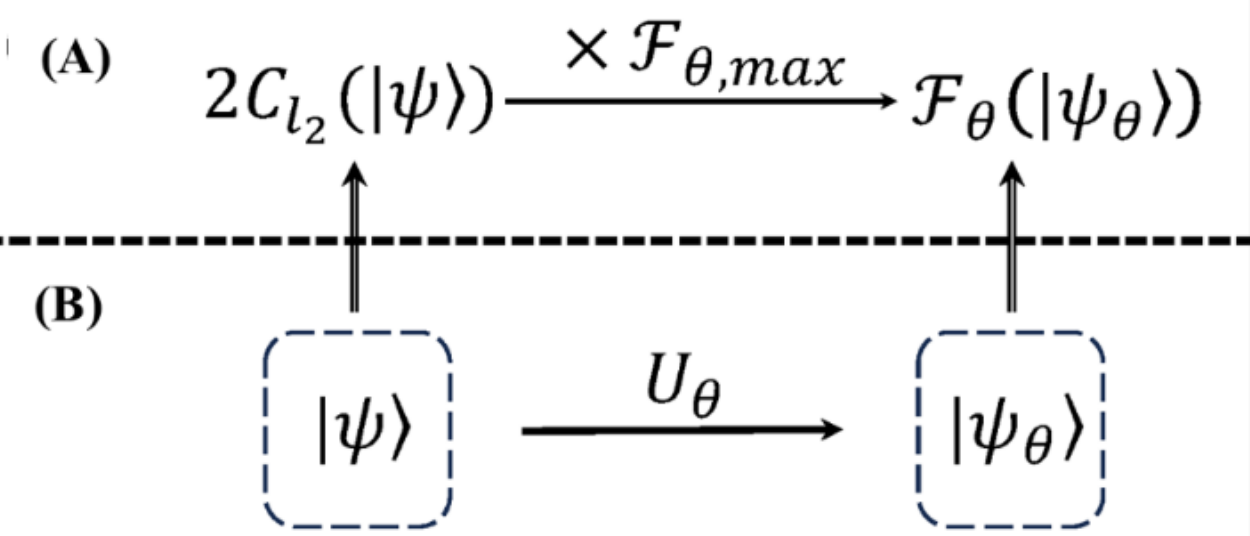}
\caption{Relation of the QFI and quantum coherence. Diagram of the factorization law for the QFI with coherence in the pure state model, $U_{\theta} = e^{- i\hat{H}\theta}$. (\textbf{A}) Factorization law for the QFI. (\textbf{B}) Corresponding evolution process.}\label{F1}
\end{figure}

\subsection*{Experimental setup}\label{sec4}

\begin{figure}[!htbp]
\centering
\includegraphics[width=1.0\linewidth]{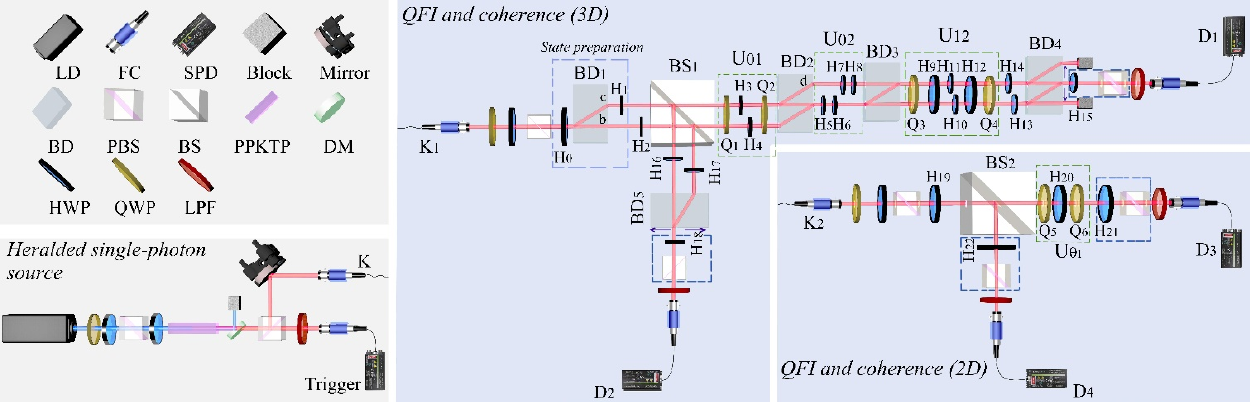}
\caption{Illustration of the experimental setup.
An experimental device for testing the factorization law for the QFI and coherence, including qubit cases (2D) and qutrit cases {[}three-dimensional (3D){]}. Pairs of single photons are generated via spontaneous parameter down-conversion (SPDC) using a type II periodically poled potassium titanyl phosphate (PPKTP) crystal. For a wave plate group consisting of an HWP and a quarter-wave plate (QWP), the signal photons with arbitrary linear polarization states are prepared by using a polarized beam splitter (PBS), beam displacer (BD), and HWPs. On a 50:50 unpolarized beam splitter (BS), the photons of the BS reflection path are measured in the initial state. The transmitted photons achieve unitary evolution through interference networks. The interference network consists of a wave plate group and BDs. The variance information was obtained by measuring $\sigma_{x}$ corresponding to the case of different qubit and qutrit cases. As shown in the figure, the photons are detected by single-photon detectors (SPDs). The purple dashed box in the figure represents the initial state preparation part in the 3D case. Because the initial state preparation in the 2D case is relatively simple and can be completed with H\textsubscript{19} alone, there is no specific annotation in the figure. The blue dashed boxes in the figure represent the measurement part for both 2D and 3D cases. Fiber couplers, FCs; long-pass filter, LPF; dichroic mirror, DM.}\label{F2}
\end{figure}

As shown in \hyperref[F2]{Fig. 2}, the main components of our setup include the initial state preparation and an interferometric network with unitary evolution operations. Last, the QFI of the final state after unitary evolution is obtained by measuring the measurement operators. For the heralded single-photon source, we use a 405-nm continuous-wave diode laser, a periodically poled potassium titanyl phosphate (PPKTP) crystal. Degenerate photon pairs produced by collinear type II spontaneous parameter down-conversion (SPDC) are collected into the fiber couplers. Because photons come in pairs, each detection of an idler photon indicates the presence of a signal photon. The K in \hyperref[F2]{Fig. 2} represents the signal photon entering the actual optical path, which is finally counted by trigger, D\textsubscript{1} and D\textsubscript{2} (D\textsubscript{3} and D\textsubscript{4}). The HWPs (H\textsubscript{0} and H\textsubscript{1}) and BD\textsubscript{1} allow a series of polarized qubit states containing path information to be encoded on the signal photon, which is then sent into the interference networks.

Moreover, we present two cases for verifying the correlation between the QFI and coherence, the qubit case, and the high-dimensional case (taking qutrit cases as an example), as shown in \hyperref[F2]{Fig. 2}. In the qubit case, K is connected to K\textsubscript{2} to input the signal photon into the subsequent optical path. The preparation of the initial state $| \psi_{2D} \rangle = \cos 2\alpha_{19} | H \rangle + \sin 2\alpha_{19} | V \rangle$ ($| H \rangle$: the horizontal polarization state; $| V \rangle$: the vertical polarization state) is relatively simple; only a polarized beam splitter (PBS) and the half-wave plate (HWP) H\textsubscript{19} can be done. Because information about the coherence of the initial state is required, BS\textsubscript{2} is used for beam splitting. The off-diagonal elements meta information of the initial state is measured by H\textsubscript{22} and PBS in the reflection path, with the photon detected by a single photon detector (D\textsubscript{4}). In the transmission path, interleaving quarter-wave plates (QWPs) Q\textsubscript{5} and Q\textsubscript{6} and H\textsubscript{20} are used to implement the $U_{\theta_{1}}$ operation. After the initial state $| \psi \rangle$ evolves under $U_{\theta}$, the final state was measured using $\sigma_{x}$ via H\textsubscript{21} and PBS, and the photons were collected with D\textsubscript{4}.

In the qutrit case, K is connected to K\textsubscript{1} to input the signal photon into the subsequent optical path. In the qutrit case, the initial state preparation is different from the qubit case, and the path information needs to be encoded in the polarization state. The prepared initial state is $| \psi_{3D} \rangle = a_{0} | 0 \rangle + a_{1} | 1 \rangle + a_{2} | 2 \rangle = \sin 2\alpha_{1}\sin 2\alpha_{0} | 0 \rangle - \cos 2\alpha_{1}\sin 2\alpha_{0} | 1 \rangle +\cos 2\alpha_{0}{ | 2 \rangle}$, where $| 0 \rangle := | H \rangle | c \rangle , | 1 \rangle := | V \rangle | c \rangle , | 2 \rangle := | H \rangle | b \rangle$, and $\sum_{i = 0}^{2}{| a_{i} |}^{2} = 1$. The path b represents the path through which the light completely passes, while path c refers to the path that is shifted upward on the basis of the original light path. We used PBS, HWPs H\textsubscript{0} and H\textsubscript{1}, and BD\textsubscript{1} for the initial state preparation (H\textsubscript{2} function is only to complete the subsequent interference networks, which does not affect the initial state preparation), as shown in \hyperref[F2]{Fig. 2}. In addition, we need information about the coherence of the initial state; here, BS\textsubscript{1} is used for beam splitting. Because the polarization information is divided into two paths by BD\textsubscript{1}, the two paths (b and c) are combined into one path by using wave plates H\textsubscript{16}, H\textsubscript{17}, and BD\textsubscript{5} in the BS\textsubscript{1} reflection path. To obtain the complete information of the initial state, it is necessary to measure with H\textsubscript{18}, PBS, and moving BD\textsubscript{5}, and the photons is detected by D\textsubscript{2}. The path contains polarization information; we decompose the unitary operation by the established methods (see the Supplementary Materials for details), i.e., $U_{\theta_{2}} = U_{12}U_{02}U_{01}$, where $U_{\mathit{ij}}$ is the unitary operations acting on the 2D subspaces of the qutrit system (\emph{\ref{R55}}), with the complementary subspace unchanged (see the Supplementary Materials for details). Wave plates H\textsubscript{3--12} and Q\textsubscript{1--4} were used to prepare the unitary operator $U_{\mathit{ij}}$, and the final state was measured by $\sigma_{x}$ using H\textsubscript{15}, BD\textsubscript{4}, and PBS; the photons were collected with D\textsubscript{1}.

\subsection*{Experimental data and results}\label{sec5}

\begin{figure}[!htbp]
\centering
\includegraphics[width=0.94\linewidth]{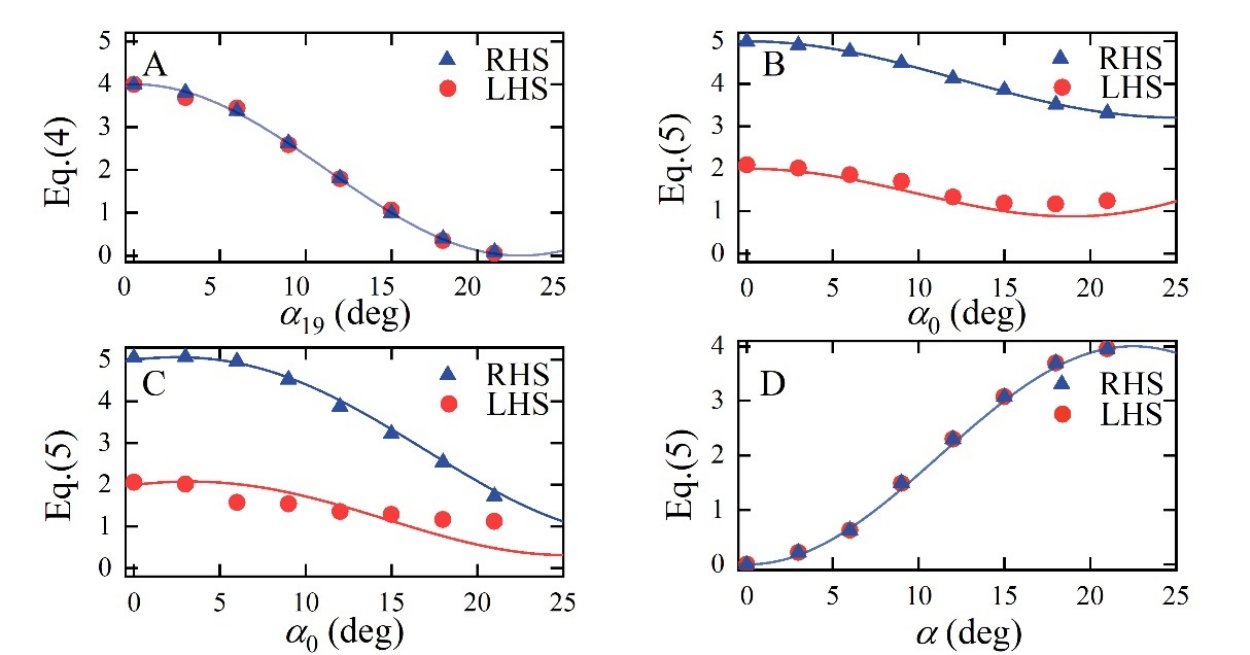}
\caption{Results for the QFI and quantum coherence.
Factorization law for the QFI and coherence in the pure state, including (\textbf{A}) qubit cases and (\textbf{B} to \textbf{D}) qutrit cases. Using a fixed combination of unitary operators, we input different initial states. We denote $\alpha_{k}$ as the angle of the HWP H\emph{\textsubscript{k}}, $\alpha_{19}$ as the angle of the initial states for qubit cases, and $\alpha_{0}$ ($\alpha$) as the angle of the initial states for qutrit cases ($\alpha$ is prepared jointly by $\alpha_{0}$ and $\alpha_{1}$). (A) \hyperref[eq:4]{Equation 4} is used to verify the factorization law of the QFI and coherence in qubit cases. The blue and red dots display the experimental values for the right-hand side ($\text{RHS} = 2C_{l_{2}}\mathcal{F}_{\theta_{1},\text{max}}$) and the left-hand side ($\text{LHS} = \mathcal{F}_{\theta_{1}}$) in \hyperref[eq:4]{Eq. 4}, and the theoretical value curve is represented by the solid line. (B to D) \hyperref[eq:5]{Equation 5} is used to verify the factorization law of the QFI and coherence in qutrit cases (3D). The experimental values for the $\text{RHS} = 2C_{l_{2}}\mathcal{F}_{\theta_{2},\text{max}}$ and $\text{LHS} = \mathcal{F}_{\theta_{2}}$ of \hyperref[eq:5]{Eq. 5} are indicated by blue and red dots, respectively, while the theoretical values are depicted by the solid lines. Panels (B) and (C) present the result diagrams for $\alpha_{1}$ set at $0{^\circ}$ and $10{^\circ}$, respectively, with each diagram resulting from adjustments to the angle $\alpha_{0}$. (D) The condition for the equality to hold in \hyperref[eq:5]{Eq. 5} is that the initial state must be a superposition state of $| 0_{\text{eig}} \rangle$ and $| 2_{\text{eig}} \rangle$ in the eigenstates of ${\hat{H}}_{2}$, expressed as $| \psi' \rangle = \sin 2\alpha | 0_{\text{eig}} \rangle + \cos 2\alpha | 2_{\text{eig}} \rangle$ (see the Supplementary Materials for details). The error bars for the 3$\sigma$ deviations from the Poisson distribution are smaller than the size of the points and have been omitted.}\label{F3}
\end{figure}

In the following, we denote $\alpha_{k}$ as the angle of the HWP H\emph{\textsubscript{k}} (where \emph{k} varies from 0 to 22) and denote $\beta_{t}$ as the angle of the quarter wave plate Q\emph{\textsubscript{t}} (where \emph{t} varies from 1 to 6). In the qubit case, we chose the Hermitian operator ${\hat{H}}_{1}$ as ${\hat{H}}_{1} = | 0 \rangle\langle 1 | + | 1 \rangle\langle 0 |$. The corresponding unitary operation $U_{\theta_{1}}$ is shown in the Supplementary Materials. In the experiment, we choose fixed parameters and select different initial states to get different QFI. We select the parameter $\theta_{1} = 24{^\circ}$. Because the $U_{\theta_{1}}$ is implemented through the QWPs and an HWP, we choose the following angles: $\beta_{5} = 90{^\circ}$, $\beta_{6} = 90{^\circ}$, and $\alpha_{20} = 12{^\circ}$. For the measurement of $\mathcal{F}_{\theta_{1}}$, we calculate the measured value of $\mathcal{F}_{\theta_{1}}$ according to the density matrix of the measured final state. The experimental results of \hyperref[eq:4]{Eq. 4} are shown in \hyperref[F3]{Fig. 3A}, where the \emph{x} axis represents the angle $\alpha_{19}$ corresponding to the different initial state. Different initial states can be achieved by rotating the HWP H\textsubscript{19} \{$\alpha_{19} \subseteq \lbrack 0{^\circ},21{^\circ}\rbrack$\}. The maximal and minimal eigenvalues corresponding to ${\hat{H}}_{1}$ are 1 and $-1$, respectively ($\mathcal{F}_{\theta_{1},\text{max}} = 4$). Obviously, the left and right sides of \hyperref[eq:4]{Eq. 4} should be strictly equal. The unitary evolution $U_{\theta_{1}}$ in qubit cases is relatively simple, and the concrete form and results can be found in the Supplementary Materials. In our experimental results, there was a slight difference between the left and right sides. One of the sources of error was the suboptimal calibration of the wave plate, leading to a deviation from the intended initial state and the unitary operation of the construction being slightly off from the expected setup.

In the qutrit case, we chose the Hermitian operator ${\hat{H}}_{2}$ as ${\hat{H}}_{2} = ( | 0 \rangle\langle 1 | + | 1 \rangle\langle 0 | + | 1 \rangle\langle 2 | + | 2 \rangle\langle 1 | )/\sqrt{2}$. The corresponding unitary operation $U_{\theta_{2}}$ has been shown in the Supplementary Materials. The verification of factorization law of the QFI and coherence in \hyperref[eq:5]{Eq. 5} is shown in \hyperref[F3]{Fig. 3 (B to D)}. In the experiment, we choose fixed parameters and select different initial states to get different QFI. Because the selection of the initial state needs to satisfy the condition $\sum_{i < j}{| a_{i} |}^{2}{| a_{j} |}^{2} < 0.25$, we choose two cases (B and C). As shown in \hyperref[F3]{Fig. 3B}, a special case is illustrated, namely $\alpha_{1} = 0{^\circ}$. As shown in \hyperref[F3]{Fig. 3C}, a more general case is illustrated, i.e., $\alpha_{1} = 10{^\circ}$. In \hyperref[F3]{Fig. 3D}, the equality in \hyperref[eq:5]{Eq. 5} holds when the chosen initial state is a superposition state of $| 0_{\text{eig}} \rangle$ and $| 2_{\text{eig}} \rangle$, $| \psi' \rangle = \sin 2\alpha | 0_{\text{eig}} \rangle + \cos 2\alpha | 2_{\text{eig}} \rangle$, where $| 0_{\text{eig}} \rangle$ and $| 2_{\text{eig}} \rangle$ are the eigenbases of ${\hat{H}}_{2}$ (see the Supplementary Materials for details).

In addition, we select the parameter $\theta_{2} = 29{^\circ}$ in the qutrit case. We decompose the $U_{\theta_{2}}$, i.e., $U_{\theta_{2}} = U_{12}U_{02}U_{01}$. For $U_{01}$, we choose the following angles: $\beta_{1} = 0{^\circ}$, $\beta_{2} = 0{^\circ}$, $\alpha_{3} = - 10{^\circ}$, and $\alpha_{4} = 0{^\circ}$. The H\textsubscript{2} function is only to complete the subsequent interference networks, so $\alpha_{2} = 45{^\circ}$. For $U_{02}$, we choose the following angles: $\alpha_{5} = 2{^\circ}$ and $\alpha_{6} = 0{^\circ}$. The H\textsubscript{7} function is only to complete the subsequent interference networks, so $\alpha_{7} = 45{^\circ}$. The H\textsubscript{8} function is as a compensating wave plate, so $\alpha_{8} = 0{^\circ}$. For $U_{12}$, we choose the following angles: $\beta_{3} = 0{^\circ}$, $\beta_{4} = 0{^\circ}$, $\alpha_{9} = 0{^\circ}$, $\alpha_{10} = 0{^\circ}$, $\alpha_{11} = 10{^\circ}$, and $\alpha_{12} = 0{^\circ}$. For the unitary evolution $U_{\theta_{2}}$ in the case of qutrit, when $\theta_{2} = 29{^\circ}$, there is a slight error between the corresponding theoretical matrix and the matrix prepared in the experiment, and the concrete form and results can be found in the Supplementary Materials. For the measurement of $\mathcal{F}_{\theta_{2}}$, we calculate the measured value of $\mathcal{F}_{\theta_{2}}$ according to the density matrix of the measured final state. The maximal and minimal eigenvalues corresponding to ${\hat{H}}_{2}$ are 1 and $-1$, respectively ($\mathcal{F}_{\theta_{2},\text{max}} = 4$). According to the theoretical result, the left side of \hyperref[eq:5]{Eq. 5} should be less than or equal to the right side. As shown in \hyperref[F3]{Fig. 3}, both cases corresponding to (B) $\alpha_{1} = 0{^\circ}$ and (C) $\alpha_{1} = 10{^\circ}$ satisfy \hyperref[eq:5]{Eq. 5}, where the \emph{x} axis represents the angles $\alpha_{0}$ corresponding to the different initial states. Obviously, in \hyperref[F3]{Fig. 3 (B and C)}, the experimental values corresponding to \hyperref[eq:5]{Eq. 5} are relatively compatible with the theoretical curve. In \hyperref[F3]{Fig. 3D}, we verify the equality of the right-hand side (RHS) and left-hand side (LHS) of \hyperref[eq:5]{Eq. 5}, where the \emph{x} axis represents the angles of $\alpha$ corresponding to different initial states. It should be emphasized that the preparation of $\alpha$ is accomplished through the combination of H\textsubscript{0} ($\alpha_{0}$) and H\textsubscript{1} ($\alpha_{1}$) (see the Supplementary Materials for details). The possible sources of error include the imperfections of the prepared pure state, the imperfect separation of the horizontal and vertical polarizations by the BDs, and the deviation of the experimentally prepared unitary operators from the theoretical one.

Our experiments further validated a lower bound on the variance of the estimated parameter on the basis of the relationship between $\text{Var}( \hat{\theta} )$ and $1/\mathcal{F}_{\theta}( | \psi_{\theta} \rangle )$, with the results depicted in \hyperref[F4]{Fig. 4}. An arbitrary Hermitian operator $\hat{A}$ (see the Supplementary Materials for details) is used as a measurement operator to estimate the parameter $\theta$ by measuring the normalized final state $| \psi_{\theta} \rangle$ postevolution and using the maximum likelihood estimation (see the Supplementary Materials for details). In the 2D scenario, the relationship between $\text{Var}( {\hat{\theta}}_{1} ) \times n$ and $1/\mathcal{F}_{\theta_{1}}{( | \psi_{\theta_{1}} \rangle)}$ is illustrated in \hyperref[F4]{Fig. 4}, where the \emph{x} axis represents the angles $\alpha_{19}$ corresponding to the different initial states. For the processing of the final state, we selected the projection measurement operator $|0\rangle\langle 0|$ on the Pauli $\sigma_{z}$ basis. The experimental results indicate that the estimated $\text{Var}( {\hat{\theta}}_{1} ) \times n$ fits the theoretical curve quite well, and our findings also reveal that when the initial state is $| \psi_{0} \rangle = \cos 2\alpha_{19} | 0 \rangle + \sin 2\alpha_{19} | 1 \rangle = | 0 \rangle$, where $\alpha_{19} = 0{^\circ}$, using $|0\rangle\langle 0|$ as the measurement operator can achieve optimal measurement (see the Supplementary Materials for details). In this work, we introduce a method for quantum parameter estimation in quantum systems by decomposing the estimation process into components related to the initial state and the quantum channel, as in \hyperref[eq:6]{Eq. 6}. This method reveals that the accuracy of quantum parameter estimation is influenced by both the initial state and the quantum channel properties. Notably, in complex scenarios, measuring only the coherence of the initial state is sufficient for accurate estimation. Certainly, one can optimize the initial state to minimize estimation variance, thereby enhancing precision.

\begin{figure}[t]
\centering
\includegraphics[width=0.86\linewidth]{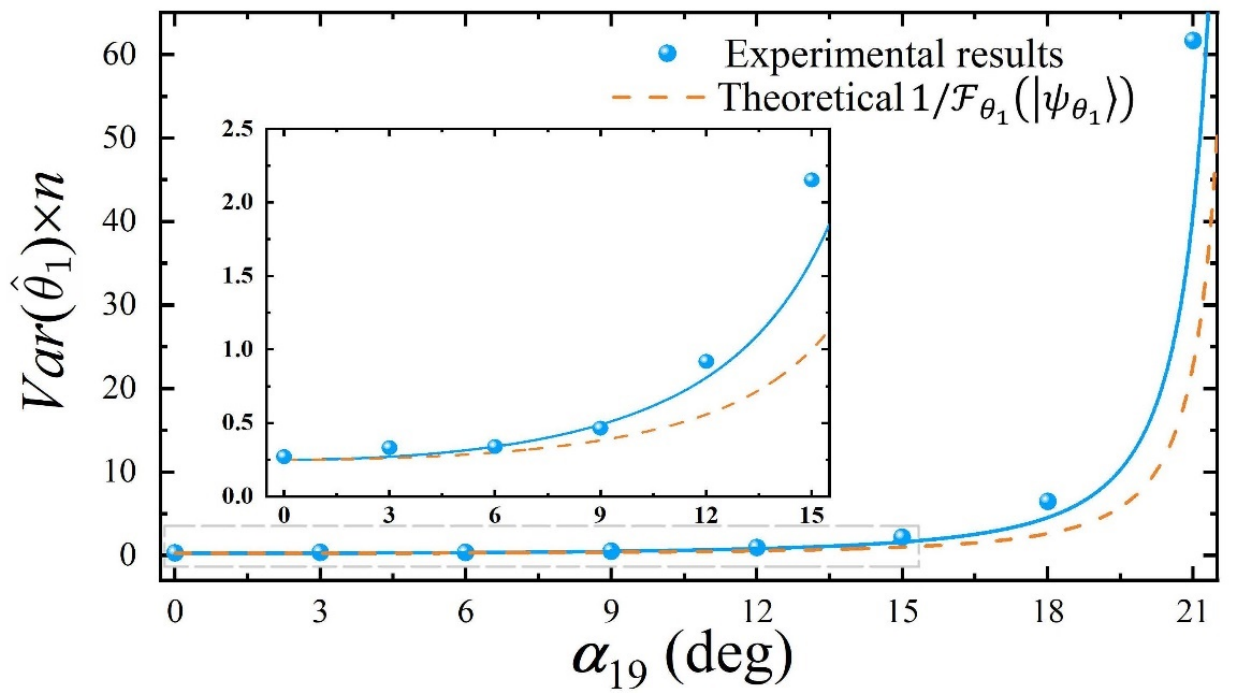}
\caption{$\textbf{Var}( \hat{\mathbf{\theta}} )$ and $1/\mathcal{F}_{\mathbf{\theta}}( | \mathbf{\psi}_{\mathbf{\theta}} \rangle )$ for varying the initial states of qubit cases.
\hyperref[eq:6]{Equation 6} is used to verify the relation of $\text{Var}( {\hat{\theta}}_{1} ) \times n$ and $1/\mathcal{F}_{\theta_{1}}{( | \psi_{\theta_{1}} \rangle)}$ in qubit cases. The experimental variances (blue dots) roughly correspond to the theoretical estimation accuracies (blue solid lines) (see the Supplementary Materials for details). The theoretical curve of $1/\mathcal{F}_{\theta_{1}}{( | \psi_{\theta_{1}} \rangle)}$ is represented by an orange dashed line.}\label{F4}
\end{figure}

\section*{DISCUSSION}\label{sec6}

We propose a measurement strategy for quantum parameter estimation that diverges from conventional methodologies. When direct measurement of the QFI ($\mathcal{F}_{\theta}$) is infeasible, the coherence of the initial state serves as a criterion for establishing the lower bound of quantum parameter estimation (\hyperref[eq:6]{Eq. 6}; see the Supplementary Materials for details). In addition, we have identified several other works (\emph{\ref{R56}}--\emph{\ref{R58}}) that warrant further exploration and citation by researchers in the field. Moreover, some works on non-Hermitian quantum metrology have demonstrated remarkable performance, achieving (\emph{\ref{R59}}) or even surpassing the Heisenberg limit under certain chosen conditions (\emph{\ref{R60}}). Compared to the previous works, our theoretical framework is not confined to qubit scenarios but has been expanded to include the more general case of qudit, thereby broadening the applicability of our findings. Furthermore, we provide a perspective that overcomes the conventional reliance on final-state information for obtaining the QFI. Our proposed factorization law enhances the lower bound on parameter estimation variance by optimizing the coherence of the input state and the quantum channel. This comprehensive approach has been substantiated through experimental validation.

In this work, the factorization law for pure states was theoretically and experimentally investigated. However, only the qubit and qutrit cases were tested. If one considers higher dimensions with $D>3$, the experiment will be more challenging, which will be our future research. In addition, we have provided a detailed discussion of the factorization law of the QFI in mixed states (see the Supplementary Materials for details).

\section*{MATERIALS AND METHODS}\label{sec7}

\subsection*{Factorization law of the QFI for pure states}\label{sec8}

First, let us consider the qubit case ($D=2$). We define the initial state $| \psi \rangle$ and the $\hat{H}$ as follows

\begin{equation}
\widehat{H} = \sum_{i = 0}^{1}{\lambda_{i}\left| i \right\rangle\left\langle i \right|},\ \ |\psi\rangle = \sum_{i = 0}^{1}{a_{i}\left| i \right\rangle}
\label{eq:7}
\end{equation}

It is worth noting that the basis for the initial state $| \psi \rangle$ needs to be expanded under the eigenstates of $\hat{H}$. Then, the corresponding QFI is $\mathcal{F}_{\theta}( | \psi_{\theta} \rangle ) = 4( \lambda_{1} - \lambda_{0} )^{2}{| a_{0} |}^{2}{| a_{1} |}^{2}$. We stipulate $\lambda_{1} > \lambda_{0}$, and \hyperref[eq:2]{Eq. 2} can be rewritten as $\mathcal{F}_{\theta,\text{max}} = ( \lambda_{1} - \lambda_{0} )^{2}$. In the qubit case, on the basis of \hyperref[eq:3]{Eq. 3}, one can obtain $C_{l_{2}}( | \psi \rangle ) = 2{| a_{0}a_{1}^{*} |}^{2}$. Therefore, we obtain the factorization law of the QFI and coherence as \hyperref[eq:4]{Eq. 4}.

Furthermore, in the qudit case where ~$D>2$, we need to first define the $| \psi \rangle$ and the $\hat{H}$ in the multidimensional case as

\begin{equation}
\widehat{H} = \sum_{i = 0}^{d - 1}{\lambda_{i}\left| i \right\rangle\left\langle i \right|},\ \ |\psi\rangle = \sum_{i = 0}^{d - 1}{a_{i}\left| i \right\rangle}
\label{eq:8}
\end{equation}

In this case, \hyperref[eq:1]{Eq. 1} is

\begin{equation}
\begin{aligned}
\mathcal{F}_{\theta}\left( \left| \psi_{\theta} \right\rangle \right) & = 4\left\lbrack \left( \sum_{i}^{}{a^{*}\left\langle i \right|} \right)\left( \sum_{j}^{}{\lambda_{j}^{2}\left| j \right\rangle\left\langle j \right|} \right)\left( \sum_{k}^{}{a_{k}\left| k \right\rangle} \right) \right\rbrack \\
 & - 4\left\lbrack \left( \sum_{i}^{}{a^{*}\left\langle i \right|} \right)\left( \sum_{j}^{}{\lambda_{j}\left| j \right\rangle\left\langle j \right|} \right)\left( \sum_{k}^{}{a_{k}\left| k \right\rangle} \right) \right\rbrack^{2} \\
 & = 4\left\lbrack \sum_{i}^{}\lambda_{i}^{2}|a_{i}|^{2} - (\sum_{i}^{}\lambda_{i}|a_{i}|^{2})^{2} \right\rbrack
\end{aligned}
\label{eq:9}
\end{equation}

using $\sum_{i}{| a_{i} |}^{2} = 1$, \hyperref[eq:9]{Eq. 9} becomes to

\begin{equation}
\begin{aligned}
\mathcal{F}_{\theta}\left( \left| \psi_{\theta} \right\rangle \right) & = 4\left\lbrack \left( \sum_{i}^{}\lambda_{i}^{2}|a_{i}|^{2} \right)\left( \sum_{j}^{}{|a_{j}|^{2}} \right) - (\sum_{i}^{}\lambda_{i}|a_{i}|^{2})^{2} \right\rbrack \\
 & = 4\left\lbrack \sum_{i < j}^{}{\left( \lambda_{i} - \lambda_{j} \right)^{2}|a_{i}|^{2}|a_{j}|^{2}} \right\rbrack
\end{aligned}
\label{eq:10}
\end{equation}

For the coherence of states in the qudit case, on the basis of \hyperref[eq:3]{Eq. 3}, one can obtain $C_{l_{2}}{( | \psi \rangle)} = 2\sum_{i < j}{| a_{i} |}^{2}{| a_{j} |}^{2}$.

Hence, in the qudit case, the factorization law of the QFI and coherence can be obtained as \hyperref[eq:5]{Eq.5}, it should be noted here that the selection of the initial state must meet the condition of $\sum_{i < j}{| a_{i} |}^{2}{| a_{j} |}^{2} < 0.25$, and \hyperref[eq:5]{Eq. 5} becomes trivial when $\sum_{i < j}{| a_{i} |}^{2}{| a_{j} |}^{2} \geq 0.25$.

\subsection*{Experimental design}\label{sec9}

Our experiments were designed to investigate the relationship between the QFI and quantum coherence, with additional discussions on quantum parameter estimation. By adjusting different wave plate angles, we quantify their association with the desired unitary evolution operations. Measurements on the final state after evolution are conducted to calculate the QFI, quantum coherence, and the variance of the estimated parameters for different initial states. All experimental data are accurate and genuine.

\subsection*{Statistical analysis}\label{sec10}

To mitigate experimental errors arising from variations in coincidence event rates over extended periods, we implemented a strategy of recording coincidence events within a 1-s window. Furthermore, we adjusted the initial state angle every 180 data points. This methodology effectively reduced the variability in measurement counts between two distinct projection operators.

\newpage

\setcounter{equation}{0}
\setcounter{figure}{0}
\renewcommand{\theequation}{S\arabic{equation}}
\renewcommand{\thefigure}{S\arabic{figure}}
\renewcommand{\thetable}{S\arabic{table}}

\section*{Supplementary Materials}\label{sec14}

\section*{A. The factorization law of QFI for pure states}

In the pure states, the Hamiltonian quantity $\hat{H}$ is independent of the parameter $\theta$. Then the quantum state $|\psi_{\theta}\rangle$ with parametric information is $|\psi_{\theta}\rangle=U_{\theta}|\psi\rangle=e^{-i\hat{H}\theta}|\psi\rangle$, and the QFI with parameter information is the form of the following variance
\begin{align}\label{SR1}
\mathcal{F}_{\theta}(|\psi_\theta\rangle)
= 4[\langle\psi_{\theta}|\hat{H}^2|\psi_{\theta}\rangle-\langle\psi_{\theta}|\hat{H}|\psi_{\theta}\rangle^2].
\end{align}
The QFI is maximised when the input state is a  superposition of eigenstates corresponding to maximal and minimal eigenvalues, a situation called channel QFI, which can be obtained as (\emph{\ref{R12}})
\begin{align}\label{SR2}
\mathcal{F}_{\theta,max}=(\lambda_{max}-\lambda_{min})^2,
\end{align}
$\lambda_{max}$ and $\lambda_{min}$ correspond to the maximum and minimum values of the eigenvalues of $\hat{H}$.
Coherence is an important physical resource, so we can use it to get the expression form of QFI that includes state coherence. In theory, the coherence of a quantum state can be measured by measuring its off-diagonal elements, the $l_2$-norm of coherence is defined as (\emph{\ref{R54}})
\begin{align}\label{SR3}
C_{l_2}(\rho)=\sum_{\substack{i,j\\i\neq j}}|\rho_{i,j}|^2.
\end{align}

Firstly, let us consider the qubit case ($d=2$). We define the initial state $|\psi\rangle$ and the $\hat{H}$ as follows
\begin{align}
\hat{H}=\sum_{i=0}^1\lambda_{i}|i\rangle\langle i|,\qquad|\psi\rangle=\sum_{i=0}^1a_{i}|i\rangle.\nonumber
\end{align}
It is worth noting that the basis for the initial state $|\psi\rangle$ needs to be expanded under the eigenstates of $\hat{H}$.
Then the corresponding QFI is
\begin{align}\label{SR4}
\mathcal{F}_{\theta}(|\psi_{\theta}\rangle)=4(\lambda_1-\lambda_0)^2|a_{0}|^2|a_{1}|^2.
\end{align}
We stipulate $\lambda_{1}>\lambda_{0}$, Eq.~(\ref{SR2}) can be rewritten as $\mathcal{F}_{\theta,max}=(\lambda_{1}-\lambda_{0})^2$.
In the case of qubit, based on Eq.~(\ref{SR3}) one can obtain $C_{l_2}(|\psi\rangle)=2|a_{0}a^*_{1}|^2$.
Therefore, we obtain the factorization law of QFI and coherence as
\begin{align}\label{SR7}
\mathcal{F}_{\theta}(|\psi_{\theta}\rangle)=2C_{l_2}(|\psi\rangle)\mathcal{F}_{\theta,max}.
\end{align}

Furthermore, in the qudit case where $d>2$, we need to first define the $|\psi\rangle$ and the $\hat{H}$ in the multidimensional case as
\begin{align}
\hat{H}=\sum_{i=0}^{d-1}\lambda_{i}|i\rangle\langle i|,\qquad|\psi\rangle=\sum_{i=0}^{d-1}a_{i}|i\rangle.\nonumber
\end{align}
In this case, Eq.~(\ref{SR1}) is
\begin{align}\label{SR8}
\mathcal{F}_{\theta}(|\psi_{\theta}\rangle)=&4[(\sum_{i}a^*\langle i|)(\sum_{j}\lambda_{j}^2|j\rangle\langle j|)(\sum_{k}a_{k}|k\rangle)] \nonumber\\ &-4[(\sum_{i}a^*\langle i|)(\sum_{j}\lambda_{j}|j\rangle\langle j|)(\sum_{k}a_{k}|k\rangle)]^2 \nonumber\\ =&4[\sum_{i}\lambda_{i}^2|a_{i}|^2-(\sum_{i}\lambda_{i}|a_{i}|^2)^2],
\end{align}
using $\sum_{i}|a_{i}|^2=1$, the Eq.~(\ref{SR8}) becomes to
\begin{align}\label{SR9}
\mathcal{F}_{\theta}(|\psi_{\theta}\rangle)&=4[(\sum_{i}\lambda_{i}^2|a_{i}|^2)(\sum_{j}|a_{j}|^2)-(\sum_{i}\lambda_{i}|a_{i}|^2)^2] \nonumber\\ &=4[\sum_{i<j}(\lambda_{i}-\lambda_{j})^2|a_{i}|^2|a_{j}|^2].
\end{align}
For the coherence of states in qudit case, based on Eq.~(\ref{SR3}) one can obtain $C_{l_2}(|\psi\rangle)=2\sum_{i<j}|a_{i}|^2|a_{j}|^2$.

Hence in the qudit case,  the factorization law of QFI and coherence can be obtained as
\begin{align}\label{SR11}
\mathcal{F}_{\theta}(|\psi_{\theta}\rangle)\leq2C_{l_2}(|\psi\rangle)\mathcal{F}_{\theta,max},
\end{align}
it should be noted here that the selection of the initial state must meet the condition of $\sum_{i<j}|a_{i}|^2|a_{j}|^2<0.25$, and Eq. (\ref{SR11}) becomes trivial when $\sum_{i<j}|a_{i}|^2|a_{j}|^2\geq0.25$.

According to the QCRB,
\begin{align}\label{ADD1}
\emph{Var}(\hat{\theta})\geq\frac{1}{n\mathcal{F}(\theta)},
\end{align}
$n$ denotes the number of measurements,
$\hat{\theta}$ is said to be an unbiased estimator of $\theta$, $\mathcal{F}(\theta)$ refers to the QFI of final states. Based on Eqs.~(\ref{SR11}) and~(\ref{ADD1}) we can get
\begin{align}\label{ADD2}
\emph{Var}(\hat{\theta})\geq\frac{1}{n\mathcal{F}_{\theta}(|\psi_{\theta}\rangle)}\geq\frac{1}{2nC_{l_2}(|\psi\rangle)\mathcal{F}_{\theta,max}},
\end{align}
 we obtain a lower bound on the variance of the parameter estimate.\\

\section*{B. The factorization law of QFI for mixed states}

In this section, we primarily derive the factorization law of QFI in mixed states:\\

\textbf{Case.I} The factorization law of QFI still holds for single qubit ($d=2$) mixed states.\\

\textbf{Proof.} To prove
\begin{align}
\mathcal{F}(\rho_{\theta})=2C_{l_{2}}(\rho_0)\mathcal{F}_{max},
\label{II.1}
\end{align}
$\mathcal{F}(\rho_{\theta})$ is associated with the chosen Hamiltonian $H$.
The left hand side (LHS) of equation (\ref{II.1}) can be written as (\emph{\ref{R8}},\emph{\ref{R9}}, \emph{\ref{R11}}, \emph{\ref{R15}}, \emph{\ref{R23}}, \emph{\ref{R61}}) 
\begin{align}
\textmd{LHS}=\mathcal{F}(\rho_{\theta})=\mathcal{F}(\rho_{0})=2\sum_{i,j}\frac{(\kappa_{i}-\kappa_j)^2}{\kappa_i+\kappa_j}|\langle\varphi_i|H|\varphi_j\rangle|^2,
\label{II.2}
\end{align}
where $\{\kappa_{i},|\varphi_{i}\rangle\}$ are eigenvalues and corresponding eigenstates of $\rho_{0}$.

When $d=2$, the initial mixed state $\rho_0$ is written as
\begin{align}
\rho_0=\kappa_1|\varphi_1\rangle\langle\varphi_1|+\kappa_2|\varphi_2\rangle\langle\varphi_2|,
\label{II.3}
\end{align}
with $\kappa_1+\kappa_2=1$ and $\langle\varphi_i|\varphi_j\rangle=\delta_{ij}$. Thus, we have
\begin{align}
\textmd{LHS}=\mathcal{F}(\rho_{0})
=&4(\kappa_{1}-\kappa_2)^2|\langle\varphi_1|H|\varphi_2\rangle|^2\nonumber\\=&4(\kappa_{1}-\kappa_2)^2\langle\varphi_1|H|\varphi_2\rangle\langle\varphi_2|H|\varphi_1\rangle\nonumber\\=&4(\kappa_{1}-\kappa_2)^2\langle\varphi_1|H(\mathbb{I}-|\varphi_1\rangle\langle\varphi_1|)H|\varphi_1\rangle\nonumber\\=&4(\kappa_{1}-\kappa_2)^2(\langle\varphi_1|H^2|\varphi_1\rangle-\langle\varphi_1|H|\varphi_1\rangle^2)\nonumber\\=&(\kappa_{1}-\kappa_2)^22C_{l_2}(|\varphi_1\rangle)\mathcal{F}_{max}.\label{II.4}
\end{align}

Let us now consider the right hand (RHS)  of equation (\ref{II.1}), we stipulate $\lambda_1>\lambda_0$
\begin{align}
\mathcal{F}_{max}=(\lambda_{1}-\lambda_{0})^2,
\label{II.5}
\end{align}
with $\lambda_{1}$ and $\lambda_{0}$ being the maximal and minimal eigenvalues of $H$. Under the basis of eigenstates of $H$ is $\{|i\rangle\}$, we have
\begin{align}
&H=\lambda_{0}|0\rangle\langle 0|+\lambda_{1}|1\rangle\langle 1|,\\
&|\varphi_1\rangle=\cos{\frac{\theta}{2}}|0\rangle+e^{i\epsilon}\sin{\frac{\theta}{2}}|1\rangle,\\
&|\varphi_2\rangle=\sin{\frac{\theta}{2}}|0\rangle-e^{i\epsilon}\cos{\frac{\theta}{2}}|1\rangle.
\label{II.6}
\end{align}
The coherence $C_{l_2}(|\varphi_1\rangle)$ and $C_{l_2}(\rho_0)$ can be written as
\begin{align}
&C_{l_2}(|\varphi_1\rangle)=\frac{1}{2}|\sin{\theta}|^2,\\
&C_{l_2}(\rho_0)=\frac{1}{2}(\kappa_1-\kappa_2)^2|\sin{\theta}|^2.
\label{II.7}
\end{align}
Thus,
\begin{align}
&\textmd{LHS}=(\kappa_{1}-\kappa_2)^22C_{l_2}(|\varphi_1\rangle)\mathcal{F}_{max}=(\kappa_{1}-\kappa_2)^2|\sin{\theta}|^2(\lambda_{1}-\lambda_{0})^2,\\
&\textmd{RHS}=2C_{l_{2}}(\rho_0)\mathcal{F}_{max}=(\kappa_{1}-\kappa_2)^2|\sin{\theta}|^2(\lambda_{1}-\lambda_{0})^2.
\label{II.8}
\end{align}
Finally, we get
\begin{align}
\textmd{LHS}=\textmd{RHS}.
\label{II.9}
\end{align}

\textbf{Case.II} Let us introduce the convex roof of $l_2$-norm coherence for mixed states ($d>2$),
\begin{align}
\widetilde{C}_{l_{2}}(\rho)=\inf_{\{p_i,|\zeta_i\rangle\}}\sum_{i}p_{i}C_{l_2}(|\zeta_i\rangle),
\label{II.10}
\end{align}
where the infimum is taken all over possible pure state decompositions of $\rho=\sum_{i}p_{i}|\zeta_{i}\rangle\langle\zeta_{i}|$. This convex roof construction has been widely used in entanglement and coherence theory. It is worth noticing that $\widetilde{C}_{l_{2}}(|\zeta\rangle)=C_{l_2}(|\zeta\rangle)$ according to Eq.~(\ref{II.10}).

Therefore, one can generalize the factorization law of QFI for pure states into the law for mixed states,
\begin{align}
\mathcal{F}_{\theta}(\rho_{\theta})\leq2\widetilde{C}_{l_{2}}(\rho_0)\mathcal{F}_{max}.
\label{II.11}
\end{align}
To prove Eq.~(\ref{II.11}), let us start from RHS. Suppose that $\rho_0=\sum_{j}q_{j}|\eta_{j}\rangle\langle\eta_{j}|$ is the optimal decomposition for $\rho_0$ to achieve the infimum of $\widetilde{C}_{l_{2}}(\rho)$ in Eq.~(\ref{II.10}). Thus,
\begin{align}
2\widetilde{C}_{l_{2}}(\rho_0)\mathcal{F}_{max}&=\sum_{j}q_{j}2C_{l_2}(|\eta_{j}\rangle)\mathcal{F}_{max}\nonumber\\&\geq\sum_{j}q_{j}\mathcal{F}_{\theta}(|\eta_{j}\rangle)\nonumber\\&\geq\mathcal{F}_{\theta}(\rho_{0})=\mathcal{F}_{\theta}(\rho_{\theta}),
\label{II.12}
\end{align}
where for the first inequality we have used the factorization law of QFI for pure states $\{|\eta_{j}\rangle\}$, and for the second inequality we have used the convex property of QFI.

In summary, we have completed the derivation of the factorization law for QFI in mixed states.\\

\section*{C. The scheme of the experiment}

Before detailing our experimental procedure, we first introduce the important components in the optical components:

$1).$ We use half-wave plates (HWP: $\alpha$) and quarter-wave plates (QWP: $\beta$) to implement unitary operation. The $\alpha$ or $\beta$ here refers to the angle between the fast axis of the waveplate and the horizontal polarization direction. The $Jones$ $matrix$ of waveplates can be denoted as (\emph{\ref{R62}})
\begin{eqnarray}
HWP(\alpha)=
   \begin{pmatrix}
     \cos2\alpha&\sin2\alpha\\
     \sin2\alpha&-\cos2\alpha
   \end{pmatrix},
QWP(\beta)=
   \begin{pmatrix}
     \cos^2\beta+i\sin^2\beta&(1-i)\cos\beta\sin\beta\\
     (1-i)\cos\beta\sin\beta&\sin^2\beta+i\cos^2\beta
   \end{pmatrix}
\end{eqnarray}

\begin{table*}
	\centering
	\caption{The preparation of the unitary operators $U_{\theta_{1}}$ in the case of 2D.}
\resizebox{\linewidth}{!}{
	\begin{tabular}{cccc}
     \hline
     \hline
		State prepared & Result  \\	
   \hline  
$H_{19}$ & $\cos2\alpha_{19}|0\rangle+\sin2\alpha_{19}|1\rangle$ \\
$Q_{5}$ $(\beta_{5}=90^\circ)$ & $i\cos2\alpha_{19}|0\rangle+\sin2\alpha_{19}|1\rangle$ \\
$H_{20}$ & $(i\cos2\alpha_{19}\cos\alpha_{20}+\sin\alpha_{19}\sin\alpha_{20})|0\rangle+(i\cos2\alpha_{19}\sin\alpha_{20}-\sin\alpha_{19}\cos\alpha_{20})|1\rangle$\\
$Q_{6}$ $(\beta_{6}=90^\circ)$ & $(-\cos2\alpha_{19}\cos\alpha_{20}+i\sin\alpha_{19}\sin\alpha_{20})|0\rangle+(i\cos2\alpha_{19}\sin\alpha_{20}-\sin\alpha_{19}\cos\alpha_{20})|1\rangle$\\
     \hline
     \hline
	\end{tabular}
                  }
    \label{tab1}
\end{table*}

$2).$ We have replaced the horizontal polarization state $|H\rangle$ and the vertical polarization state $|V\rangle$ as follows
\begin{eqnarray}
|H\rangle=|0\rangle=
   \begin{pmatrix}
    1\\
     0
   \end{pmatrix},
|V\rangle=|1\rangle=
   \begin{pmatrix}
    0\\
     1
   \end{pmatrix}
\end{eqnarray}

$3).$ The beam displacer (BD) is capable of fully transmitting horizontally polarized photons, but diverting them from their original path when they emit vertically polarized photons ($4mm$ in our experiments).

\begin{figure}[b]
\includegraphics[width=9cm]{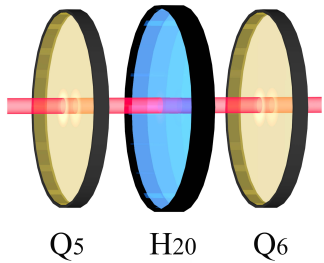}
\caption{Experimental setup of unitary operators $U_{\theta_{1}}$ in the case of 2D.}\label{FigII}
\end{figure}

\subsection*{Preparation of unitary operators $U_{\theta_{1}}$ and $U_{\theta_{2}}$}
An experimental device for testing the factorization law for quantum Fisher information and coherence, including 2D cases and 3D cases.

The experimental setup of unitary operator $U_{\theta_{1}}$ in 2D is shown in Fig.~\ref{FigII}. We choose the following angles $\beta_{5} = 90^\circ$, $\beta_{6} = 90^\circ$, $\alpha_{20}$, so the matrix form of $U_{\theta_{1}}$ be expressed as
\begin{eqnarray}
U_{\theta_{1}}=-
   \begin{pmatrix}
     \cos2\alpha_{20} & -i\sin2\alpha_{20}\\
     -i\sin2\alpha_{20} & \cos2\alpha_{20}
   \end{pmatrix},
\end{eqnarray}
$\theta_{1}=2\alpha_{20}$. In practice, we select the parameter $\theta_{1}=24^\circ$. The experimentally derived $U_{\theta_{1}}$ differs from the theoretically calculated matrix by an overall negative sign. However, this does not impact the following experimental results. In the 2D case, Table~\ref{tab1} illustrates the relevant processes.

\begin{table*}[t!]
	\centering
	\caption{The preparation of the unitary operators $U_{01}$ in the case of 3D.}
\resizebox{\linewidth}{!}{
	\begin{tabular}{cccc}
     \hline
     \hline
		State prepared & Result  \\	
   \hline  
$H_{0}$ & $\cos2\alpha_{0}|0\rangle+\sin2\alpha_{0}|1\rangle$ \\
$BD_{1}$ & $\cos2\alpha_{0}|0\rangle|b\rangle+\sin2\alpha_{0}|1\rangle|c\rangle$\\
$H_{1}$ & $(\sin2\alpha_{1}\sin2\alpha_{0}|0\rangle-\cos2\alpha_{1}\sin2\alpha_{0}|1\rangle)|c\rangle$\\
$H_{2}$ $(\alpha_{2}=45^\circ)$ & $\cos2\alpha_{0}|1\rangle|b\rangle$\\
$Q_{1}$ $(\beta_{1}=0^\circ)$ & $i\cos2\alpha_{0}|1\rangle|b\rangle+(\sin2\alpha_{1}\sin2\alpha_{0}|0\rangle-i\cos2\alpha_{1}\sin2\alpha_{0}|1\rangle)|c\rangle$\\
$H_{3}$ & $((\cos2\alpha_{3}\sin2\alpha_{1}\sin2\alpha_{0}-i\sin2\alpha_{3}\cos2\alpha_{1}\sin2\alpha_{0})|0\rangle+(i\cos2\alpha_{3}\cos2\alpha_{1}\sin2\alpha_{0}+\sin2\alpha_{3}\sin2\alpha_{1}\sin2\alpha_{0})|1\rangle)|c\rangle$\\
$H_{4}$ $(\alpha_{4}=0^\circ)$ & $-i\cos2\alpha_{0}|1\rangle|b\rangle$\\
$Q_{2}$ $(\beta_{2}=0^\circ)$ & $\cos2\alpha_{0}|1\rangle|b\rangle+((\cos2\alpha_{3}\sin2\alpha_{1}\sin2\alpha_{0}-i\sin2\alpha_{3}\cos2\alpha_{1}\sin2\alpha_{0})|0\rangle+(-\cos2\alpha_{3}\cos2\alpha_{1}\sin2\alpha_{0}+i\sin2\alpha_{3}\sin2\alpha_{1}\sin2\alpha_{0})|1\rangle)|c\rangle$\\
     \hline
     \hline
	\end{tabular}
}
    \label{tab2}
\end{table*}

\begin{table*}[t]
	\centering
	\caption{The preparation of the unitary operators $U_{02}$ in the case of 3D}
\resizebox{\linewidth}{!}{
	\begin{tabular}{{cp{15cm}c}}
     \hline
     \hline
		State prepared & Result  \\	
   \hline  
$BD_{2}$ & $(\cos2\alpha_{0}|1\rangle+(\cos2\alpha_{3}\sin2\alpha_{1}\sin2\alpha_{0}-i\sin2\alpha_{3}\cos2\alpha_{1}\sin2\alpha_{0})|0\rangle)|c\rangle+(-\cos2\alpha_{3}\cos2\alpha_{1}\sin2\alpha_{0}+i\sin2\alpha_{3}\sin2\alpha_{1}\sin2\alpha_{0})|1\rangle|d\rangle$\\
$H_{5}$ & $(\sin2\alpha_{5}\cos2\alpha_{0}+\cos2\alpha_{5}(\cos2\alpha_{3}\sin2\alpha_{1}\sin2\alpha_{0}-i\sin2\alpha_{3}\cos2\alpha_{1}\sin2\alpha_{0}))|0\rangle|c\rangle+(-\cos2\alpha_{5}\cos2\alpha_{0}+\sin2\alpha_{5}(\cos2\alpha_{3}\sin2\alpha_{1}\sin2\alpha_{0}-i\sin2\alpha_{3}\cos2\alpha_{1}\sin2\alpha_{0}))|1\rangle|c\rangle$\\
$H_{6}$ $(\alpha_{6}=0^\circ)$ & $(\sin2\alpha_{5}\cos2\alpha_{0}+\cos2\alpha_{5}(\cos2\alpha_{3}\sin2\alpha_{1}\sin2\alpha_{0}-i\sin2\alpha_{3}\cos2\alpha_{1}\sin2\alpha_{0}))|0\rangle|c\rangle+(\cos2\alpha_{5}\cos2\alpha_{0}-\sin2\alpha_{5}(\cos2\alpha_{3}\sin2\alpha_{1}\sin2\alpha_{0}-i\sin2\alpha_{3}\cos2\alpha_{1}\sin2\alpha_{0}))|1\rangle|c\rangle$\\
$H_{7}$ $(\alpha_{7}=45^\circ)$ & $(-\cos2\alpha_{3}\cos2\alpha_{1}\sin2\alpha_{0}+i\sin2\alpha_{3}\sin2\alpha_{1}\sin2\alpha_{0})|0\rangle|d\rangle$\\
$H_{8}$ $(\alpha_{8}=0^\circ)$ & $(-\cos2\alpha_{3}\cos2\alpha_{1}\sin2\alpha_{0}+i\sin2\alpha_{3}\sin2\alpha_{1}\sin2\alpha_{0})|0\rangle|d\rangle$\\
     \hline
     \hline
	\end{tabular}
}
    \label{tab3}
\end{table*}

\begin{table*}[t]
	\centering
	\caption{The preparation of the unitary operators $U_{12}$ in the case of 3D}
\resizebox{\linewidth}{!}{
	\begin{tabular}{cp{16cm}c}
     \hline
     \hline
		State prepared & Result  \\	
   \hline  
$BD_{3}$ & $(\sin2\alpha_{5}\cos2\alpha_{0}+\cos2\alpha_{5}(\cos2\alpha_{3}\sin2\alpha_{1}\sin2\alpha_{0}-i\sin2\alpha_{3}\cos2\alpha_{1}\sin2\alpha_{0}))|0\rangle|c\rangle+((\cos2\alpha_{5}\cos2\alpha_{0}-\sin2\alpha_{5}(\cos2\alpha_{3}\sin2\alpha_{1}\sin2\alpha_{0}-i\sin2\alpha_{3}\cos2\alpha_{1}\sin2\alpha_{0}))|1\rangle+(-\cos2\alpha_{3}\cos2\alpha_{1}\sin2\alpha_{0}+i\sin2\alpha_{3}\sin2\alpha_{1}\sin2\alpha_{0})|0\rangle)|d\rangle$\\
$Q_{3}$ $(\beta_{3}=0^\circ)$ & $(\sin2\alpha_{5}\cos2\alpha_{0}+\cos2\alpha_{5}(\cos2\alpha_{3}\sin2\alpha_{1}\sin2\alpha_{0}-i\sin2\alpha_{3}\cos2\alpha_{1}\sin2\alpha_{0}))|0\rangle|c\rangle+(i(-\cos2\alpha_{5}\cos2\alpha_{0}-\sin2\alpha_{5}(\cos2\alpha_{3}\sin2\alpha_{1}\sin2\alpha_{0}-i\sin2\alpha_{3}\cos2\alpha_{1}\sin2\alpha_{0}))|1\rangle+(-\cos2\alpha_{3}\cos2\alpha_{1}\sin2\alpha_{0}+i\sin2\alpha_{3}\sin2\alpha_{1}\sin2\alpha_{0})|0\rangle)|d\rangle$\\
$H_{9}$ $(\alpha_{9}=0^\circ)$ & $(\sin2\alpha_{5}\cos2\alpha_{0}+\cos2\alpha_{5}(\cos2\alpha_{3}\sin2\alpha_{1}\sin2\alpha_{0}-i\sin2\alpha_{3}\cos2\alpha_{1}\sin2\alpha_{0}))|0\rangle|c\rangle+(-i(-\cos2\alpha_{5}\cos2\alpha_{0}-\sin2\alpha_{5}(\cos2\alpha_{3}\sin2\alpha_{1}\sin2\alpha_{0}-i\sin2\alpha_{3}\cos2\alpha_{1}\sin2\alpha_{0}))|1\rangle+(-\cos2\alpha_{3}\cos2\alpha_{1}\sin2\alpha_{0}+i\sin2\alpha_{3}\sin2\alpha_{1}\sin2\alpha_{0})|0\rangle)|d\rangle$\\
$H_{10}$ $(\alpha_{10}=0^\circ)$ & $(\sin2\alpha_{5}\cos2\alpha_{0}+\cos2\alpha_{5}(\cos2\alpha_{3}\sin2\alpha_{1}\sin2\alpha_{0}-i\sin2\alpha_{3}\cos2\alpha_{1}\sin2\alpha_{0}))|0\rangle|c\rangle$\\
$H_{11}$ & $(\cos\alpha_{11}(-\cos2\alpha_{3}\cos2\alpha_{1}\sin2\alpha_{0}+i\sin2\alpha_{3}\sin2\alpha_{1}\sin2\alpha_{0})-i\sin\alpha_{11}(-\cos2\alpha_{5}\cos2\alpha_{0}-\sin2\alpha_{5}(\cos2\alpha_{3}\sin2\alpha_{1}\sin2\alpha_{0}-i\sin2\alpha_{3}\cos2\alpha_{1}\sin2\alpha_{0})))|0\rangle|d\rangle+(\sin\alpha_{11}(-\cos2\alpha_{3}\cos2\alpha_{1}\sin2\alpha_{0}+i\sin2\alpha_{3}\sin2\alpha_{1}\sin2\alpha_{0})+i\cos\alpha_{11}(-\cos2\alpha_{5}\cos2\alpha_{0}-\sin2\alpha_{5}(\cos2\alpha_{3}\sin2\alpha_{1}\sin2\alpha_{0}-i\sin2\alpha_{3}\cos2\alpha_{1}\sin2\alpha_{0})))|1\rangle|d\rangle$\\
$H_{12}$ $(\alpha_{12}=0^\circ)$ &
$(\cos\alpha_{11}(-\cos2\alpha_{3}\cos2\alpha_{1}\sin2\alpha_{0}+i\sin2\alpha_{3}\sin2\alpha_{1}\sin2\alpha_{0})-i\sin\alpha_{11}(-\cos2\alpha_{5}\cos2\alpha_{0}-\sin2\alpha_{5}(\cos2\alpha_{3}\sin2\alpha_{1}\sin2\alpha_{0}-i\sin2\alpha_{3}\cos2\alpha_{1}\sin2\alpha_{0})))|0\rangle|d\rangle+(-\sin\alpha_{11}(-\cos2\alpha_{3}\cos2\alpha_{1}\sin2\alpha_{0}+i\sin2\alpha_{3}\sin2\alpha_{1}\sin2\alpha_{0})-i\cos\alpha_{11}(-\cos2\alpha_{5}\cos2\alpha_{0}-\sin2\alpha_{5}(\cos2\alpha_{3}\sin2\alpha_{1}\sin2\alpha_{0}-i\sin2\alpha_{3}\cos2\alpha_{1}\sin2\alpha_{0})))|1\rangle|d\rangle+(\sin2\alpha_{5}\cos2\alpha_{0}+\cos2\alpha_{5}(\cos2\alpha_{3}\sin2\alpha_{1}\sin2\alpha_{0}-i\sin2\alpha_{3}\cos2\alpha_{1}\sin2\alpha_{0}))|0\rangle|c\rangle$\\
$Q_{4}$ $(\beta_{4}=0^\circ)$ &
$(\cos\alpha_{11}(-\cos2\alpha_{3}\cos2\alpha_{1}\sin2\alpha_{0}+i\sin2\alpha_{3}\sin2\alpha_{1}\sin2\alpha_{0})-i\sin\alpha_{11}(-\cos2\alpha_{5}\cos2\alpha_{0}-\sin2\alpha_{5}(\cos2\alpha_{3}\sin2\alpha_{1}\sin2\alpha_{0}-i\sin2\alpha_{3}\cos2\alpha_{1}\sin2\alpha_{0})))|0\rangle|d\rangle+(-i\sin\alpha_{11}(-\cos2\alpha_{3}\cos2\alpha_{1}\sin2\alpha_{0}+i\sin2\alpha_{3}\sin2\alpha_{1}\sin2\alpha_{0})+\cos\alpha_{11}(-\cos2\alpha_{5}\cos2\alpha_{0}-\sin2\alpha_{5}(\cos2\alpha_{3}\sin2\alpha_{1}\sin2\alpha_{0}-i\sin2\alpha_{3}\cos2\alpha_{1}\sin2\alpha_{0})))|1\rangle|d\rangle+(\sin2\alpha_{5}\cos2\alpha_{0}+\cos2\alpha_{5}(\cos2\alpha_{3}\sin2\alpha_{1}\sin2\alpha_{0}-i\sin2\alpha_{3}\cos2\alpha_{1}\sin2\alpha_{0}))|0\rangle|c\rangle$\\
     \hline
     \hline
	\end{tabular}
}
    \label{tab4}
\end{table*}

In the 3D case, we decompose the unitary operation by the established methods in (\emph{\ref{R55}}), i.e., $U_{\theta_{2}}=U_{12}U_{02}U_{01}$, where $U_{ij}$ is the unitary operations acting on the 2D subspaces of the qutrit system, with the complementary subspace unchanged. The $U_{\theta_{2}}$ can be written as
\begin{eqnarray}
U_{\theta_{2}}=
   \begin{pmatrix}
     \frac{1}{2}(1+\cos\theta_{2}) & -\frac{i\sin\theta_{2}}{\sqrt{2}} & \frac{1}{2}(-1+\cos\theta_{2})\\
     -\frac{i\sin\theta_{2}}{\sqrt{2}} & \cos\theta_{2} & -\frac{i\sin\theta_{2}}{\sqrt{2}}\\
     \frac{1}{2}(-1+\cos\theta_{1}) & -\frac{i\sin\theta_{2}}{\sqrt{2}} & \frac{1}{2}(1+\cos\theta_{2})\\
   \end{pmatrix}.
\end{eqnarray}

\begin{figure*}[b]
\includegraphics[width=16.5cm]{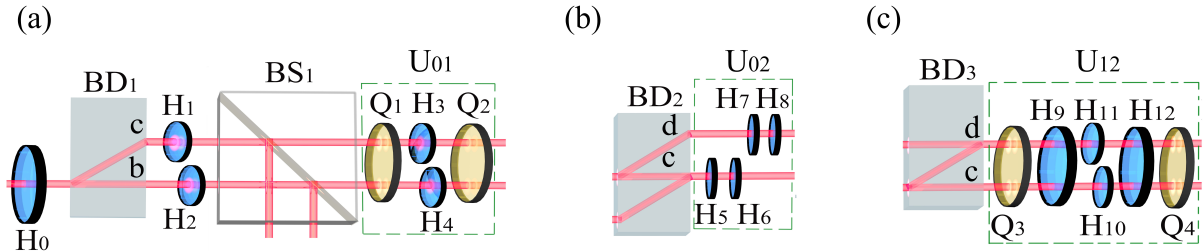}
\caption{Experimental setup of unitary operators $U_{\theta_{2}}$ in the case of 3D. (a) $U_{01}$. (b) $U_{02}$. (c) $U_{12}$.} \label{FigIII}
\end{figure*}

According to the theoretical decomposition, the decomposed $U_{01}$, $U_{02}$ and $U_{12}$ takes the following theoretical for as
\begin{eqnarray}
U_{01}&=&
   \begin{pmatrix}
     \frac{1+\cos\theta_{2}}{\sqrt{-(-3+\cos\theta_{2})(1+\cos\theta_{2})}} & \frac{2i\sin\theta_{2}}{\sqrt{5+4\cos\theta_{2}-\cos2\theta_{2}}} & 0\\
     \frac{2i\sin\theta_{2}}{\sqrt{5+4\cos\theta_{2}-\cos2\theta_{2}}} & \frac{1+\cos\theta_{2}}{\sqrt{-(-3+\cos\theta_{2})(1+\cos\theta_{2})}} & 0\\
     0 & 0 & 1\\
   \end{pmatrix},\\
U_{02}&=&
   \begin{pmatrix}
     \frac{1}{2}\sqrt{(3-\cos\theta_{2})(1+\cos\theta_{2})} & 0 & \frac{1}{2}(-1+\cos\theta_{2})\\
     0 & 1 & 0\\
     \frac{1}{2}(1-\cos\theta_{2}) & 0 & \frac{1}{2}\sqrt{(3-\cos\theta_{2})(1+\cos\theta_{2})}\\
   \end{pmatrix},\\
U_{12}&=&
   \begin{pmatrix}
     1 & 0 & 0\\
     0 & \frac{1+\cos\theta_{2}}{3-\cos\theta_{2}} & -\frac{i\sqrt{2}\sin\theta_{2}}{\sqrt{(3-\cos\theta_{2})(1+\cos\theta_{2})}}\\
     0 & -\frac{i\sqrt{2}\sin\theta_{2}}{\sqrt{(3-\cos\theta_{2})(1+\cos\theta_{2})}} & \frac{1+\cos\theta_{2}}{3-\cos\theta_{2}}\\
   \end{pmatrix}.
\end{eqnarray}
The experimental setup of unitary operator $U_{01}$, $U_{02}$ and $U_{12}$ in 3D is shown in Fig.~\ref{FigIII}. We choose the following angles $\beta_{1}=0^\circ$, $\beta_{2}=0^\circ$ and $\alpha_{4}=0^\circ$ for $U_{12}$. The function of $U_{01}$ is to implement unitary operation on the $c$ path and unit operation on the $b$ path, as shown in Table~\ref{tab2}. We choose $\theta_{2}=29^\circ$, so $\alpha_{3}=-10^\circ$.

We choose the following angles $\alpha_{6}=0^\circ$, $\alpha_{7}=45^\circ$ and $\alpha_{8}=0^\circ$ for $U_{02}$. The function of $U_{02}$ is to implement unitary operation on the $d$ path and unit operation on the $c$ path, as shown in Table~\ref{tab3}. We choose $\theta_{2}=29^\circ$, so $\alpha_{5}=2^\circ$.

We choose the following angles $\alpha_{9}=0^\circ$, $\alpha_{10}=0^\circ$, $\alpha_{12}=0^\circ$, $\beta_{3}=0^\circ$ and $\beta_{4}=0^\circ$ for $U_{12}$. The function of $U_{12}$ is to implement unitary operation on the $c$ path and unit operation on the $d$ path, as shown in Table~\ref{tab4}. We choose $\theta_{2}=29^\circ$, so $\alpha_{11}=10^\circ$.

\begin{table*}[t]
	\centering
	\caption{Experimental results for the 2D case regarding the value of $2C_{l_2}\mathcal{F}_{\theta_{1},max}$ and $\mathcal{F}_{\theta_1}(|\psi_{\theta_1}\rangle)$}
    \resizebox{\linewidth}{!}{
	\begin{tabular}{ccccc}
     \hline
     \hline
      $\alpha_{19}$ & $2C_{l_2}\mathcal{F}_{\theta_{1},max}$ & Theoretical value & $\mathcal{F}_{\theta_1}(|\psi_{\theta_1}\rangle)$ & Theoretical value \\
   \hline  
$0^\circ$ & $3.985$ & 4.000 & $3.993$ & 4.000 \\
$3^\circ$ & $3.811$ & 3.827 & $3.689$ & 3.827 \\
$6^\circ$ & $3.376$ & 3.338 & $3.438$ & 3.338 \\
$9^\circ$ & $2.624$ & 2.618 & $2.585$ & 2.618 \\
$12^\circ$ & $1.807$ & 1.791 & $1.797$ & 1.791 \\
$15^\circ$ & $0.996$ & 1.000 & $1.057$ & 1.000 \\
$18^\circ$ & $0.402$ & 0.382 & $0.354$ & 0.382 \\
$21^\circ$ & $0.091$ & 0.044 & $0.048$ & 0.044\\
    \hline
    \hline
	\end{tabular}
}
    \label{tab5}
\end{table*}
\begin{table*}[t]
    \centering
	\caption{Experimental results for the 3D case regarding the value of $2C_{l_2}\mathcal{F}_{\theta_{2},max}$ and $\mathcal{F}_{\theta_2}(|\psi_{\theta_2}\rangle)$}
   \resizebox{\linewidth}{!}{
	\begin{tabular}{ccccccccccccccccc}
     \hline
     \hline
      $\alpha_{0} (\alpha_{1}=0^\circ)$ & $2C_{l_2}\mathcal{F}_{\theta_{2},max}$ & Theoretical value & $\mathcal{F}_{\theta_2}(|\psi_{\theta_2}\rangle)$ & Theoretical value & $\alpha_{0} (\alpha_{1}=10^\circ)$ & $2C_{l_2}\mathcal{F}_{\theta_{2},max}$ & Theoretical value & $\mathcal{F}_{\theta_2}(|\psi_{\theta_2}\rangle)$ & Theoretical value\\
   \hline  
$0^\circ$ & $4.994$ & 5.000 & $2.088$ & 2.000 &$0^\circ$ & $5.050$ & 5.000 & $2.062$ & 2.000 \\
$3^\circ$ & $4.912$ & 4.935 & $2.007$ & 1.935 &$3^\circ$ & $5.070$ & 5.060 & $2.019$ & 2.080 \\
$6^\circ$ & $4.759$ & 4.750 & $1.853$ & 1.756 &$6^\circ$ & $4.965$ & 4.924 & $1.581$ & 2.018 \\
$9^\circ$ & $4.491$ & 4.473 & $1.697$ & 1.500 &$9^\circ$ & $4.526$ & 4.573 & $1.545$ & 1.817 \\
$12^\circ$ & $4.123$ & 4.144 & $1.333$ & 1.226 &$12^\circ$ & $3.879$ & 4.025 & $1.361$ & 1.506 \\
$15^\circ$ & $3.851$ & 3.813 & $1.180$ & 1.000 &$15^\circ$ & $3.229$ & 3.333 & $1.289$ & 1.134 \\
$18^\circ$ & $3.512$ & 3.524 & $1.163$ & 0.882 &$18^\circ$ & $2.546$ & 2.579 & $1.169$ & 0.771 \\
$21^\circ$ & $3.313$ & 3.316 & $1.247$ & 0.917 &$21^\circ$ & $1.731$ & 1.860 & $1.126$ & 0.483 \\
    \hline
    \hline
	\end{tabular}
}\label{tab6}
\end{table*}

\subsection*{Experimental value}
Because we choose $\sigma_{x}$ as the Hermitian operator $\hat{H}_{1}$, so $\mathcal{F}_{\theta_1}(|\psi_{\theta_1}\rangle)$ become as
\begin{equation}
\mathcal{F}_{\theta_1}(|\psi_{\theta_1}\rangle)=4[\langle\psi_{\theta_1}|\sigma_{x}^2|\psi_{\theta_1}\rangle-\langle\psi_{\theta_1}|\sigma_{x}|\psi_{\theta_1}\rangle^2]=4(1-\langle\psi_{\theta_1}|\sigma_{x}|\psi_{\theta_1}\rangle^2).  \label{UR2}
\end{equation}
We listed the experimental data on the left and right sides of Eq.~(\ref{SR7}) in the 2D case. As shown in Table~\ref{tab5}.

In the 3D case, the actorization law of QFI and coherence as follows
\begin{equation}
\mathcal{F}_{\theta_2}(|\psi_{\theta_2}\rangle)\leq2C_{l_2}(|\psi\rangle)\mathcal{F}_{\theta_2,max}.  \label{UR3}
\end{equation}
We choose $S_{x}$ as the Hermitian operator $\hat{H}_{2}$, so $\mathcal{F}_{\theta_2}(|\psi\rangle)$ become as
\begin{equation}
\mathcal{F}_{\theta_2}(|\psi_{\theta_2}\rangle)=4[\langle\psi_{\theta_2}|S_{x}^2|\psi_{\theta_2}\rangle-\langle\psi_{\theta_2}|S_{x}|\psi_{\theta_2}\rangle^2].  \label{UR4}
\end{equation}
Because the selection of the initial state needs to satisfy the condition $\sum_{i<j}|a_{i}|^2|a_{j}|^2\leq0.25$, we choose two cases $\alpha_{1}=0^\circ$ and $\alpha_{1}=10^\circ$. The experimental data of Eq.~(\ref{SR11}) as shown in Table~\ref{tab6}.\\

\subsection*{A tight bound for Eq.~(5) in the main text ($d=3$)}
Taking the 3D case as an example, the left and right sides of Eq.~(5) in the main text are strictly equal when the initial state chosen is exactly the state with the largest superposition of the eigenstates of the selected Hamiltonian $\hat{H}_{2}=(|0\rangle\langle1|+|1\rangle\langle0|+|1\rangle\langle2|+|2\rangle\langle1|)/\sqrt{2}$. The eigenvalues $-1$, $0$, $1$ of $\hat{H}_{2}$ correspond to the eigenbases represented by

\begin{align}
|\rm{0_{eig}}\rangle&=\frac{1}{2}|0\rangle-\frac{1}{\sqrt{2}}|1\rangle+\frac{1}{2}|2\rangle,\nonumber\\
|\rm{1_{eig}}\rangle&=-\frac{1}{\sqrt{2}}|0\rangle+\frac{1}{\sqrt{2}}|2\rangle,\nonumber\\
|\rm{2_{eig}}\rangle&=\frac{1}{2}|0\rangle+\frac{1}{\sqrt{2}}|1\rangle+\frac{1}{2}|2\rangle. \label{B1}
\end{align}

Obviously, the theoretically required initial state is the superposition of states $|\rm{0_{eig}}\rangle$ and $|\rm{2_{eig}}\rangle$, with their corresponding coefficients only needing to satisfy $|\lambda_{0}^\prime|^2+|\lambda_{2}^\prime|^2=1$. For convenience, the  initial state $|\psi\rangle$ can be written as

\begin{align}
|\psi\rangle&=\lambda_{0}^\prime|\rm{0_{eig}}\rangle+\lambda_{2}^\prime|\rm{2_{eig}}\rangle,  \label{S1}
\end{align}

Based on Eq.~(\ref{S1}), the coherence be written as

\begin{equation}
C_{l_2}(|\psi\rangle)=2|\lambda^\prime_{0}\lambda^\prime_{2}|^2.  \label{S2}
\end{equation}

Then the corresponding QFI is
\begin{align}
\mathcal{F}_{\theta_{2}}(|\psi_{\theta_2}\rangle)=4(\langle \psi_{\theta_2}|\hat{H}_2^2|\psi_{\theta_2}\rangle-\langle \psi_{\theta_2}|\hat{H}_2|\psi_{\theta_2}\rangle^2)=16|\lambda^\prime_{0}\lambda^\prime_{2}|^2, \label{F1}
\end{align}

Finally, we can establish the relationship between the coherence and QFI, expressed as
\begin{equation}
\mathcal{F}_{\theta_{2}}(|\psi_{\theta_2}\rangle)=2C_{l_2}(|\psi\rangle)\mathcal{F}_{\theta_{2},max}, \label{E1}
\end{equation}
where $\mathcal{F}_{\theta_{2},max}=4$.

Based on Eq.~(\ref{S1}), $\lambda_{0}^\prime=\sin{2\alpha}$ and $\lambda_{2}^\prime=\cos{2\alpha}$. The computational basis as

\begin{align}
|\phi^\prime\rangle&=\sin{2\alpha}|\rm{0_{eig}}\rangle+\cos{2\alpha}|\rm{2_{eig}}\rangle\nonumber\\&=\frac{1}{2}(\cos2\alpha+\sin2\alpha)|0\rangle+\frac{\cos2\alpha-\sin2\alpha}{\sqrt{2}}|1\rangle+\frac{1}{2}(\cos2\alpha+\sin2\alpha)|2\rangle \label{S3}
\end{align}
In the actual experimental process, we need to combine the half-wave plates $H_{0}$ and $H_{1}$ to prepare the computational basis required for the initial state that conforms to Eq.~(\ref{S3}). Then perform the corresponding calculations based on the measurement results. The prepared initial state $|\phi^\prime\rangle$ is defined as
\begin{align}
|\phi^{\prime\prime}\rangle&=\lambda_{0}^{\prime\prime}|0\rangle+\lambda_{1}^{\prime\prime}|1\rangle+\lambda_{2}^{\prime\prime}|2\rangle\nonumber\\&=\sin{2\alpha_{1}}\sin{2\alpha_{0}}|0\rangle-\cos{\alpha_{1}}\sin{2\alpha_{0}}|1\rangle+\cos{2\alpha_{0}}|2\rangle. \label{S4}
\end{align}

Based on Eq.~(\ref{S3}) and Eq.~(\ref{S4}), we can get the different values of $\alpha_{0}$ and $\alpha_{1}$ from each $\alpha$ ($\alpha\subseteq [0^\circ, 21^\circ]$).

The left and right sides of Eq.~(\ref{E1}) should be strictly equal. The experimental data of Eq.~(\ref{E1}) as shown in Table~\ref{tab7}.

\begin{table*}[t]
	\centering
	\caption{Experimental results for the 3D case regarding the value of $2C_{l_2}\mathcal{F}_{\theta_{2},max}$ and $\mathcal{F}_{\theta_2}(|\psi_{\theta_2}\rangle)$}
    \resizebox{1.0\linewidth}{!}{
	\begin{tabular}{cccccccccccccccccccccc}
     \hline
     \hline
      $\alpha$ & $\alpha_{0}$ & Theoretical value & $\alpha_{1}$ & Theoretical value & $2C_{l_2}\mathcal{F}_{\theta_{2},max}$ & Theoretical value & $\mathcal{F}_{\theta_2}(|\psi_{\theta_2}\rangle)$ & Theoretical value\\
   \hline  
$0^\circ$ & $-30^\circ$ & $-30^\circ$ & $-17.6^\circ$ & $-17.6322^\circ$ & $0.0038$ & 0.0000 & $0.0050$ & 0.0000\\
$3^\circ$ & $-28.3^\circ$ & $-28.3328^\circ$ & $-20.6^\circ$ & $-20.5638^\circ$ & $0.2192$ & 0.1729 & $0.2206$ & 0.1729 \\
$6^\circ$ & $-26.8^\circ$ & $-26.8139^\circ$ & $-23.7^\circ$ & $-23.7178^\circ$ & $0.6280$ & 0.6617 & $0.6292$ & 0.6617 \\
$9^\circ$ & $-25.5^\circ$ & $-25.4736^\circ$ & $-27.1^\circ$ & $-27.1122^\circ$ & $1.4853$ & 1.3820 & $1.4927$ & 1.3820 \\
$12^\circ$ & $-24.3^\circ$ & $-24.3447^\circ$ & $-30.8^\circ$ & $-30.752^\circ$ & $2.2975$ & 2.2091 & $2.2988$ & 2.2091 \\
$15^\circ$ & $-23.5^\circ$ & $-23.4602^\circ$ & $-34.6^\circ$ & $-34.6232^\circ$ & $3.0726$ & 3.0000 & $3.0728$ & 3.0000 \\
$18^\circ$ & $-22.9^\circ$ & $-22.8506^\circ$ & $-38.7^\circ$ & $-38.6874^\circ$ & $3.6927$ & 3.6180 & $3.6927$ & 3.6180 \\
$21^\circ$ & $-22.5^\circ$ & $-22.5392^\circ$ & $-42.9^\circ$ & $-42.8806^\circ$ & $3.9516$ & 3.9563 & $3.9535$ & 3.9563 \\
    \hline
    \hline
	\end{tabular}
} \label{tab7}
\end{table*}

\section*{D. The variance of $\theta_{1}$ for Eq.~(6) in the main text}
\subsection*{The method of the  variance of $\theta$ with $n$}
In general, to perform parameter estimation, one must select an appropriate measurement operator, which can be denoted as $\hat{A}$. The choice of $\hat{A}$ is not unique. However, to conduct the optimal measurement, it is necessary to explore all possible initial states, a process that will inevitably consume substantial resources. Therefore, the selection of a measurement operator and the identification of initial states that can achieve optimal measurement outcomes are both of considerable importance. Expanding upon the foundational experiments, the research not only investigates the relationship between the initial state's coherence and the QFI at the final state but also considers the application of measurement data derived from the final state for parameter estimation as Eq.~(\ref{ADD2}). This approach is theoretically justified and is further exemplified by the meticulous processing of relevant measurement information from the final state, specifically in scenarios where the dimension $d=2$, with the objective of performing parameter estimation under optimal measurement conditions.

To estimate the distinct parameter $\theta_{1}$, we employ the method of maximum likelihood estimation to obtain the estimator $\theta_{1}$. For the independent number of measurements $n$, the likelihood function can be expressed as

\begin{equation}
L(n,x|\theta)=\frac{n!}{(n-x)!x!}p(\theta)^x[1-p(\theta)]^{(n-x)}. \label{Methood1}
\end{equation}

In which $x$ denotes the number of measurements for the chosen measurement operator $\hat{A}$ out of $n$ total measurements, and $p(\theta)=|\langle\psi_{\theta}|\hat{A}|\psi_{\theta}\rangle|$ represents the probability of measurement under $\hat{A}$. The logarithm of the likelihood function can be written as

\begin{equation}
\ln[L(n,x|\theta)]=\ln[\frac{n!}{(n-x)!x!}]+x\ln[p(\theta)]+(n-x)\ln[1-p(\theta)], \label{Methood2}
\end{equation}
and solve the differential equation
\begin{equation}
\frac{d[L(n,x|\theta)]}{d\theta}=x\frac{d[p(\theta)]}{d\theta}+\frac{d[1-p(\theta)]}{d\theta}=0. \label{Methood3}
\end{equation}
The solution to Eq.~(\ref{Methood3}) is the estimator $\theta=\theta(n,x)$, which functions as a function of $n$ and $x$. By substituting the experimental results into $\theta(n,x)$, we obtain an estimate for the parameter $\theta$. Repeating this process allows us to further derive the distribution of the estimated binomial (\emph{\ref{R11}},\emph{\ref{R23}},\emph{\ref{R63}}-\emph{\ref{R66}}).

\subsection*{Expermental results for the variance of $\theta_{1}$ for Eq.~(6) in $d=2$}
Ensuring the tightness of the inequality on the right-hand side of Eq.~(\ref{ADD2}). Here the measurement $\hat{A}_{1}$ is denoted as
\begin{eqnarray}
\hat{A}_{1}=
   \begin{pmatrix}
     1 & 0 \\
     0 & 0 \\
   \end{pmatrix}.
\end{eqnarray}
Theoretically, the operator $\hat{A}$ is optimal for measurement only when the initial state is either $|0\rangle$ (or $|1\rangle$). Therefore, the operator
$\hat{A}$ is considered to be in an optimal configuration for the states $|0\rangle$. In our experiments, we input different initial states given by $|\phi_0\rangle=\cos{2\alpha_{19}}|0\rangle+\sin{2\alpha_{19}}|1\rangle$, where $|0\rangle$ and $|1\rangle$ expressed as the horizontal polarization state and the vertical polarization state, and use the operator $\hat{A}$ to measure the final state to verify whether the estimation accuracy of the parameter $\theta_{1}$ can only achieve the quantum Cram$\rm{\acute{e}}$r-Rao lower bound when the initial state is $|0\rangle$($\alpha_{19}=0^\circ$).

\begin{table*}[t]
	\centering
	\caption{Experimental results for the $d=2$ case regarding the value of $\emph{Var}(\hat{\theta}_1) \times n$ and $1/\mathcal{F}_{\theta_1}(|\psi_{\theta_1}\rangle)$}
    \resizebox{0.65\linewidth}{!}{
	\begin{tabular}{cccccccccccccccccccccc}
     \hline
     \hline
      $\alpha_{19}$ & $\emph{Var}(\hat{\theta}_1) \times n$ & Theoretical $\emph{Var}(\hat{\theta}_1) \times n$ & $1/\mathcal{F}_{\theta_1}(|\psi_{\theta_1}\rangle)$  \\
   \hline  
$0^\circ$ & 0.2711 & 0.2500 & 0.2500 \\
$3^\circ$ & 0.3322 & 0.2705 & 0.2613 \\
$6^\circ$ & 0.3398 & 0.3397 & 0.2996 \\
$9^\circ$ & 0.4645 & 0.4890 & 0.3820 \\
$12^\circ$ & 0.9190 & 0.8084 & 0.5584 \\
$15^\circ$ & 2.1527 & 1.6081 & 1.0000 \\
$18^\circ$ & 6.4700 & 4.5379 & 2.6180 \\
$21^\circ$ & 61.6836 & 41.2282 & 22.8811 \\
    \hline
    \hline
	\end{tabular}
} \label{add8}
\end{table*}

\begin{table*}[t]
	\centering
	\caption{Experimental results for the $d=2$ case regarding the value of $1/\mathcal{F}_{\theta_1}(|\psi_{\theta_1}\rangle)$ and $1/(2C_{l_2}\mathcal{F}_{\theta_{1},max})$ from Table~\ref{tab5}}
        \resizebox{0.65\linewidth}{!}{
	\begin{tabular}{ccccc}
     \hline
     \hline
      $\alpha_{19}$ & $[\mathcal{F}_{\theta_1}(|\psi_{\theta_1}\rangle)]^{-1}$ & Theoretical value & $[2C_{l_2}\mathcal{F}_{\theta_{1},max}]^{-1}$ & Theoretical value \\
   \hline  
$0^\circ$ & $0.2504$ & 0.2500 & $0.2509$ & 0.2500 \\
$3^\circ$ & $0.2711$ & 0.2613 & $0.2624$ & 0.2613 \\
$6^\circ$ & $0.2909$ & 0.2996 & $0.2962$ & 0.2996 \\
$9^\circ$ & $0.3868$ & 0.3820 & $0.3811$ & 0.3820 \\
$12^\circ$ & $0.5565$ & 0.5584 & $0.5534$ & 0.5584 \\
$15^\circ$ & $0.9461$ & 1.0000 & $1.0040$ & 1.0000 \\
$18^\circ$ & $2.8249$ & 2.6180 & $2.4876$ & 2.6180 \\
$21^\circ$ & $20.8333$ & 22.8811 & $10.9890$ & 22.8811\\
    \hline
    \hline
	\end{tabular}
}
    \label{tab10}
\end{table*}

Under condition $d=2$, as indicated by the experimental results presented in Table~\ref{add8}. The optimal measurement conditions are met when the initial state is $|\phi_0\rangle=|0\rangle$, which corresponds to the case when $\alpha_{19}=2k\pi$ for integer $k$. In the final state analysis, we utilized the projection measurement operator $|0\rangle\langle0|$ based on the Pauli $\sigma_z$ basis. The experimental findings demonstrate a close alignment between the calculated variance $\text{Var}(\theta_1)$ and the theoretical predictions. Moreover, our research discovered that optimal measurement outcomes are attainable when the initial state is set to $|\phi_0\rangle=|0\rangle$, which corresponds to $\alpha_{19} = 0^\circ$, by employing $|0\rangle\langle0|$ as the measurement operator.

We discussed the relationship between the parameter estimation variance and QFI when the right-hand side of Eq.~(\ref{ADD2})is tight. The corresponding results are shown in Table~\ref{add8}.
Additionally, we also considers the conditions required for optimal measurement. The corresponding theoretical curve is shown in Fig.~4 of the main text, and the fit between theory and experimental results is very good. In more general cases, Eq.~(\ref{ADD2}) also holds, but the parameter estimation process may be more complicated.

However, in a special case where the measurement scheme is incomplete, preventing the experimenter from measuring QFI, it is possible to consider using the coherence information of the initial state as the lower bound for quantum parameter estimation. Although this lower bound is not tight, there are still cases where equality holds. Based on the $2C_{l_2}\mathcal{F}_{\theta,max}$ and $\mathcal{F}_{\theta}(|\psi_{\theta}\rangle)$ from Table~\ref{tab5}, we calculated the corresponding data results for $[2C_{l_2}\mathcal{F}_{\theta,max}]^{-1}$ and $[\mathcal{F}_{\theta}(|\psi_{\theta}\rangle)]^{-1}$ as shown in Table~\ref{tab10}.
Taking $d=2$ as an example, based on the comparison between $[2C_{l_2}\mathcal{F}_{\theta_1,max}]^{-1}$ in Tables~\ref{tab10} and $Var(\theta_1)$ in Table~\ref{add8}, it is observed that the experimental data results for $[2C_{l_2}\mathcal{F}_{\theta_1,max}]^{-1}$ remain significantly promising even when the optimal measurement condition ($\alpha_{19}=0^\circ$), is satisfied.

\clearpage
\section*{REFERENCES AND NOTES}

\begin{enumerate}
\def\labelenumi{\arabic{enumi}.}
\item\label{R1} V. Giovannetti, S. Lloyd, L. Maccone, Advances in quantum metrology. Nat. Photonics 5, 222--229 (2011).

\item\label{R2} J. Dowling, Quantum optical metrology -- The lowdown on high-N00N states. Contemp. Phys. 49, 125 (2008).

\item\label{R3} J. A. Jones, S. D. Karlen, J. Fitzsimons, A. Ardavan, S. C. Benjamin, G. A. D. Briggs, J. J. L. Morton, Magnetic field sensing beyond the standard quantum limit using 10-spin NOON states. Science 324, 1330 (2009).

\item\label{R4} S. Simmons, J. A. Jones, S. D. Karlen, A. Ardavan, J. J. L. Morton, Magnetic field sensors using 13-spin cat states. Phys. Rev. A 82, 022330 (2010).

\item\label{R5} B. L. Higgins, D. W. Berry, S. D. Bartlett, H. M. Wiseman, G. J. Pryde, Entanglement-free Heisenberg-limited phase estimation. Nature 450, 393--396 (2007).

\item\label{R6} R. Demkowicz-Dobrzanski, U. Dorner, B. J. Smith, J. S. Lundeen, W. Wasilewski, K. Banaszek, I. A. Walmsley, Quantum phase estimation with lossy interferometers. Phys. Rev. A 80, 013825 (2009).

\item\label{R7} R. A. Fisher, Theory of statistical estimation. Cambridge Phil. Soc. 22, 700 (1929).

\item\label{R8} E. Martínez-Vargas, C. Pineda, F. Leyvraz, P. Barberis-Blostein, Quantum estimation of unknown parameters. Phys. Rev. A 95, 012136 (2017).

\item\label{R9} A. S. Holevo, Probabilistic and Statistical Aspects of Quantum Theory (North-Holland, 1982).

\item\label{R10} C. W. Helstrom, Quantum detection and estimation theory. J. Stat. Phys. 1, 231--252 (1969).

\item\label{R11} S. L. Braunstein, C. M. Caves, G. J. Milburn, Generalized uncertainty relations: Theory, examples and lorentz invariance. Ann. Phys. 247, 135--173 (1996).

\item\label{R12} J. Ko\l{}odyński, R. Demkowicz-Dobrzański, Efficient tools for quantum metrology with uncorrelated noise. New J. Phys. 15, 073043 (2013).

\item\label{R13} H. Cramer, Mathematical Methods of Statistics (Princeton Univ. Press, 1946).

\item\label{R14} C. W. Helstrom, Minimum mean-squared error of estimates in quantum statistics. Phys. Lett. A 25, 101--102 (1967).

\item\label{R15} G. Tóth, F. Fröwis, Uncertainty relations with the variance and the quantum Fisher information based on convex decompositions of density matrices. Phys. Rev. Res. 4, 013075 (2022).

\item\label{R16} S.-H. Chiew, M. Gessner, Improving sum uncertainty relations with the quantum Fisher information. Phys. Rev. Res. 4, 013076 (2022).

\item\label{R17} P. Hyllus, W. Laskowski, R. Krischek, C. Schwemmer, W. Wieczorek, H. Weinfurter, L. Pezzé, A. Smerzi, Fisher information and multiparticle entanglement. Phys. Rev. A 85, 022321 (2012).

\item\label{R18} G. Tóth, Multipartite entanglement and high-precision metrology. Phys. Rev. A 85, 022322 (2012).

\item\label{R19} G. Vitagliano, M. Fadel, I. Apellaniz, M. Kleinmann, B. Lücke, C. Klempt, G. Tóth, Number-phase uncertainty relations and bipartite entanglement detection in spin ensembles. Quantum 7, 914 (2023).

\item\label{R20} H. T. Song, S. L. Luo, Y. Hong, Quantum non-Markovianity based on the Fisher-information matrix. Phys. Rev. A 91, 042110 (2015).

\item\label{R21} X. M. Lu, X. Wang, C. P. Sun, Quantum Fisher information flow and non-Markovian processes of open systems. Phys. Rev. A 82, 042103 (2010).

\item\label{R22} M. Vatasescu, Dynamics of quantum Fisher information from a time-local non-Markovian master equation with decoherence rates and operators depending on the estimated parameter. Phys. Rev. A 106, 042204 (2022).

\item\label{R23} S. L. Braunstein, C. M. Caves, Statistical distance and the geometry of quantum states. Phys. Rev. Lett. 72, 3439--3443 (1994).

\item\label{R24} X. M. Lu, X. Wang, Incorporating Heisenberg’s uncertainty principle into quantum multiparameter estimation. Phys. Rev. Lett. 126, 120503 (2021).

\item\label{R25} J. Liu, H. Yuan, X. M. Lu, X. Wang, Quantum Fisher information matrix and multiparameter estimation. J. Phys. A: Math. Theor. 53, 023001 (2020).

\item\label{R26} N. Metwally, F. Ebrahim, Fisher information of accelerated two-qubit system in the presence of the color and white noisy channels. Int. J. Mod. Phys. B 34, 2050027 (2020).

\item\label{R27} X. Yu, C. Zhang, Quantum parameter estimation of non-Hermitian systems with optimal measurements. Phys. Rev. A 108, 022215 (2023).

\item\label{R28} C. Benedetti, F. Buscemi, P. Bordone, M. G. A. Paris, Quantum probes for the spectral properties of a classical environment. Phys. Rev. A 89, 032114 (2014).

\item\label{R29} A. Streltsov, G. Adesso, M. B. Plenio, Colloquium: Quantum coherence as a resource. Rev. Mod. Phys. 89, 041003 (2017).

\item\label{R30} A. Streltsov, U. Singh, H. S. Dhar, M. N. Bera, G. Adesso, Measuring quantum coherence with entanglement. Phys. Rev. Lett. 115, 020403 (2015).

\item\label{R31} Z. Ma, Z. Zhang, Y. Dai, Y. Dong, C. Zhang, Detecting and estimating coherence based on coherence witnesses. Phys. Rev. A 103, 012409 (2021).

\item\label{R32} Y.-C. Chen, J. Cheng, W.-Z. Zhang, C.-J. Zhang, Detecting coherence with respect to general quantum measurements. Sci. China Inf. Sci. 66, 180504 (2023).

\item\label{R33} M.-L. Hu, X. Hu, J. Wang, Y. Peng, Y.-R. Zhang, H. Fan, Quantum coherence and geometric quantum discord. Phys. Rep. 762-764, 1--100 (2018).

\item\label{R34} S. Pirandola, Quantum discord as a resource for quantum cryptography. Sci. Rep. 4, 6956 (2014).

\item\label{R35} E. Knill, R. Laflamme, Power of one bit of quantum information. Phys. Rev. Lett. 81, 5672--5675 (1998).

\item\label{R36} D. Pan, G.-L. Long, L. Yin, Y.-B. Sheng, D. Ruan, S. X. Ng, J. Lu, L. Hanzo, The evolution of quantum secure direct communication: On the road to the Qinternet. IEEE Commun. Surv. Tutorials 26, 3 (2024).

\item\label{R37} D. Pan, P. Niu, H. Zhang, F. Zhang, M. Wang, X.-T. Song, X. Chen, C. Zheng, G.-L. Long, Simultaneous transmission of information and key exchange using the same photonic quantum states. Sci. Adv. 11, eadt4627 (2025).

\item\label{R38} Y. Dai, Y. Dong, Z. Xu, W. You, C. Zhang, O. Gühne, Experimentally accessible lower bounds for genuine multipartite entanglement and coherence measures. Phys. Rev. Applied 13, 054022 (2020).

\item\label{R39} S. Rana, P. Parashar, M. Lewenstein, Trace-distance measure of coherence. Phys. Rev. A 93, 012110 (2016).

\item\label{R40} L. H. Shao, Z. J. Xi, H. Fan, Y. M. Li, Fidelity and trace-norm distances for quantifying coherence. Phys. Rev. A 91, 042120 (2015).

\item\label{R41} D. P. Pires, L. C. Céleri, D. O. Soares-Pinto, Geometric lower bound for a quantum coherence measure. Phys. Rev. A 91, 042330 (2015).

\item\label{R42} D. Girolami, Observable measure of quantum coherence in finite dimensional systems. Phys. Rev. Lett. 113, 170401 (2014).

\item\label{R43} E. Wolf, Coherence properties of partially polarized electromagnetic radiation. Nuovo Cim. 13, 1165--1181 (1959).

\item\label{R44} R. Barakat, N-fold polarization measures and associated thermodynamic entropy of N partially coherent pencils of radiation. Opt. Acta 30, 1171--1182 (1983).

\item\label{R45} Y. Yao, G. H. Dong, X. Xiao, C. P. Sun, Frobenius-norm-based measures of quantum coherence and asymmetry. Sci. Rep. 6, 32010 (2016).

\item\label{R46} M. A. Alonso, X.-F. Qian, J. H. Eberly, Center-of-mass interpretation for bipartite purity analysis of N-party entanglement. Phys. Rev. A 94, 030303 (2016).

\item\label{R47} A. S. M. Patoary, G. Kulkarni, A. K. Jha, Intrinsic degree of coherence of classical and quantum states. J. Opt. Soc. Am. B, Opt. Phys. 36, 2765 (2019).

\item\label{R48} K.-D. Wu, Z. Hou, Y.-Y. Zhao, G.-Y. Xiang, C.-F. Li, G.-C. Guo, J. Ma, Q.-Y. He, J. Thompson, M. Gu, Experimental cyclic interconversion between coherence and quantum correlations. Phys. Rev. Lett. 121, 050401 (2018).

\item\label{R49} C. Zhang, T. R. Bromley, Y. F. Huang, H. Cao, W. M. Lv, B. H. Liu, C. F. Li, G. C. Guo, M. Cianciaruso, G. Adesso, Demonstrating quantum coherence and metrology that is resilient to transversal noise. Phys. Rev. Lett. 123, 180504 (2019).

\item\label{R50} K.-D. Wu, T. Theurer, G.-Y. Xiang, C.-F. Li, G.-C. Guo, M. B. Plenio, A. Streltsov, Quantum coherence and state conversion: Theory and experiment. NPJ Quantum Inf. 6, 22 (2020).

\item\label{R51} Y. T. Wang, J. S. Tang, Z. Y. Wei, S. Yu, Z. J. Ke, X. Y. Xu, C. F. Li, G. C. Guo, Directly measuring the degree of quantum coherence using interference fringes. Phys. Rev. Lett. 118, 020403 (2017).

\item\label{R52} Y. Aiache, K. El Anouz, N. Metwally, A. El Allati, Dynamics of quantum coherence and nonlocality of a two-spin system in the chemical compass. Phys. Rev. E 109, 034101 (2024).

\item\label{R53} K. El Anouz, C. P. Onyenegecha, A. I. Opara, A. Salah, A. El Allati, Dynamics of quantum Fisher information and quantum coherence of two interacting atoms under time--fractional analysis. J. Opt. Soc. Am. B 39, 979--989 (2022).

\item\label{R54} T. Baumgratz, M. Cramer, M. B. Plenio, Quantifying coherence. Phys. Rev. Lett. 113, 140401 (2014).

\item\label{R55} M. Reck, A. Zeilinger, H. J. Bernstein, P. Bertani, Experimental realization of any discrete unitary operator. Phys. Rev. Lett. 73, 58--61 (1994).

\item\label{R56} J. L. Zhao, D. X. Chen, Y. Zhang, Y. L. Fang, M. Yang, Q. C. Wu, C. P. Yang, Coherence and quantum Fisher information in general single-qubit parameter estimation processes. Phys. Rev. A 104, 062608 (2021).

\item\label{R57} X. N. Feng, L. F. Wei, Quantifying quantum coherence with quantum Fisher information. Sci. Rep. 7, 15492 (2017).

\item\label{R58} D.-H. Yu, C.-S. Yu, Quantifying coherence in terms of Fisher information. Phys. Rev. A 106, 052432 (2022).

\item\label{R59} X. Yu, X. Zhao, L. Li, X.-M. Hu, X. Duan, H. Yuan, C. Zhang, Toward Heisenberg scaling in non-Hermitian metrology at the quantum regime. Sci. Adv. 10, eadk7616 (2024).

\item\label{R60} J. Li, H. Liu, Z. Wang, X. X. Yi, Enhanced parameter estimation by measurement of non-Hermitian operators. AAPPS Bull. 33, 22 (2023).

\item\label{R61} D. Petz, Quantum Information Theory and Quantum Statistics (Springer, 2008).

\item\label{R62} D. F. V. James, P. G. Kwiat, W. J. Munro, A. G. White, Measurement of qubits. Phys. Rev. A 64, 052312 (2001).

\item\label{R63} S. F. Huelga, C. Macchiavello, T. Pellizzari, A. K. Ekert, M. B. Plenio, J. I. Cirac, Improvement of frequency standards with quantum entanglement. Phys. Rev. Lett. 79, 3865--3868 (1997).

\item\label{R64} B. Yurke, S. L. McCall, J. R. Klauder, SU(2) and SU(1,1) interferometers. Phys. Rev. A 33, 4033--4054 (1986).

\item\label{R65} G. Tóth, I. Apellaniz, Quantum metrology from a quantum information science perspective. J. Phys. A Math. Theor. 47, 424006 (2014).

\item\label{R66} S. Pang, T. A. Brun, Quantum metrology for a general Hamiltonian parameter. Phys. Rev. A 90, 022117 (2014).
\end{enumerate}

\section*{Acknowledgments}\label{ack1}

\textbf{Funding:} This work is supported by the National Natural Science Foundation of China (grant no. 62475127), the Innovation Program for Quantum Science and Technology (grant no. 2021ZD0301200), the Zhejiang Provincial Natural Science Foundation of China (grant no. LZ25A040006), and K.C. Wong Magna Fund in Ningbo University.

\textbf{Author contributions:} X.Z. and C.Z. designed and performed the experiment. X.Z. and C.Z. contributed to the theoretical analysis. X.Z., X.Y., L.L., W.Z., and C.Z. analyzed the theoretical prediction and experimental data. X.Z. and C.Z. wrote the paper.

\textbf{Competing interests:} The authors declare that they have no competing interests.

\textbf{Data and materials availability:} All data needed to evaluate the conclusions in the paper are present in the paper and/or the Supplementary Materials.

\end{document}